\documentclass[fleqn,usenatbib,float]{mnras}

\usepackage{newtxtext,newtxmath}

\usepackage[T1]{fontenc}

\DeclareRobustCommand{\VAN}[3]{#2}
\let\VANthebibliography\thebibliography
\def\thebibliography{\DeclareRobustCommand{\VAN}[3]{##3}\VANthebibliography}

\usepackage{float}
\usepackage{graphicx}	
\usepackage{amsmath}	
\usepackage[nice]{nicefrac}
\usepackage{booktabs}
\usepackage{tabu}
\usepackage{siunitx}
\usepackage{makecell}
\usepackage{multirow}

\defcitealias{amorin_star_2012}{A2012}
\let\oldAA\AA
\renewcommand{\AA}{\text{\normalfont\oldAA}}

\title[Insights from GP deep spectra]{New insights on the nebular emission, ionizing radiation and low metallicity of Green Peas from advanced modelling}
\author[V. Fern\'andez]{
V. Fern\'andez,$^{1,2}$\thanks{E-mail: vital.fernandez@userena.cl}
R. Amor\'in$^{1,2}$,
E. P\'erez-Montero$^{3}$,
P. Papaderos$^{4,5}$,
C. Kehrig$^{3}$,
J. M. V\'ilchez$^{3}$
\\
$^{1}$Departamento de Astronom\'ia, Universidad de La Serena, Av. Juan Cisternas 1200 Norte, La Serena, Chile\\
$^{2}$Instituto de Investigaci\'on Multidisciplinar en Ciencia y Tecnolog\'ia, Universidad de La Serena, Ra\'ul Bitr\'an 1305, La Serena, Chile\\
$^{3}$Instituto de Astrof\'isica de Andaluc\'ia-CSIC, Glorieta de la Astronom\'ia S/N, E-18008 Granada, Spain \\
$^{4}$Centro de Astrof\'isica e Ciências do Espaço, Universidade de Lisboa - OAL, Tapada da Ajuda, PT1349-018 Lisboa, Portugal\\
$^{5}$Departamento de F\'isica, Faculdade de Ciências da Universidade de Lisboa, Edif\'cio C8, Campo Grande, PT1749-016 Lisboa, Portugal
}

\date{Accepted XXX. Received YYY; in original form ZZZ}

\pubyear{2015}

\begin{document}
\label{firstpage}
\pagerange{\pageref{firstpage}--\pageref{lastpage}}
\maketitle

\begin{abstract}
Low-metallicity, compact starburst galaxies referred to as Green Peas (GPs) provide a unique window to study galactic evolution across cosmic epochs. In this work, we present new deep optical spectra for three GPs from OSIRIS at the 10m Gran Telescopio Canarias (GTC), which are studied using a state-of-the-art methodology. A stellar population synthesis is conducted with 1098 spectral templates.  The methodology succeeds at characterising stellar populations from 0.5 Myrs to 10 Gyrs. The light distribution shows a large red excess from a single population with $log\left(age\right) > 8.5yr$ in the GP sample analysed. This points towards an incomplete characterisation of the gas luminosity, whose continuum already accounts between $7.4\%$ and $27.6\%$ in the galaxy sample. The emission spectra are fitted with the largest Bayesian chemical model consisting of a electron temperature, a electron density, the logarithmic extinction coefficient and eleven ionic species under the direct method paradigm. Additionally, building on previous work, we propose a neural networks sampler to constrain the effective temperature and ionization parameter of each source from photoionization model grids. Finally, we combine both methodologies into a 16-dimensional model, which for the first time, simultaneously explores the direct method and photoionization parameter spaces. Both techniques consistently indicate a low metallicity gas, $7.76<12+log\left(\nicefrac{O}{H}\right)<8.04$, ionized by strong radiation fields, in agreement with previous works. 

\end{abstract} 




\section{Introduction}

Understanding major phases of galaxy growth and chemical enrichment and how young galaxies have contributed to cosmic reionization are two of the most pressing questions in modern astrophysics. Current observations of high-redshift galaxies and model predictions highlight the role of low-mass starbursting galaxies as important contributors to the cosmic star formation history \citep[e.g.][]{atek_hubble_2014, maseda_muse_2018} and the reionization process \citep[e.g.][]{stark_galaxies_2016,endsley_o_2021}. These small, young systems are typically identified as extreme emission-line galaxies (EELGs) by their strong nebular emission and hard radiation fields, evidenced by the presence of high-ionization emission lines with large H$\beta$+[OIII]5007$\AA$ equivalent widths
\citep[EW $\sim$300-3000\AA][]{smit_evidence_2014,roberts-borsani_zgtrsim_2016,stark_ly_2017,barros_greats_2019}.

At lower redshift, an increasingly large number of low-mass EELGs have been discovered from $z\sim$0.5 to $z$\,$\sim$\,4 and deep spectroscopic data have allowed a global characterization of their physical  properties, which are now relatively well-established
\citep[e.g][]{wel_extreme_2011,maseda_nature_2014,amorin_evidence_2014,amorin_extreme_2015,amorin_analogues_2017,calabro_characterization_2017, forrest_discovery_2017, tang_mmtmmirs_2019, du_searching_2020,tran_mosel_2020, endsley_o_2021}. A large fraction of EELGs show high EW emission lines when observed in the rest-UV, including strong Ly$\alpha$ emission \citep[e.g.][]{erb_physical_2010,amorin_analogues_2017,berg_window_2018,tang_lyman-alpha_2021} and some of them have been detected as LyC emitters \citep[e.g.][]{barros_extreme_2016,vanzella_hubble_2016,vanzella_direct_2018}. However, a thorough characterization on these systems star formation histories and chemical abundances is still lacking. This is due to the high sensitivity, resolution and coverage spectra requirements.  These observations are still challenging in faint and compact objects -particularly in the deep NIR spectroscopy

A technical solution to the previous constrains will be available with the new generation telescopes. These include the James Webb Space Telescope (JWST) and the ground-based extremely large telescopes. However, a rare family of EELGs in the local universe can still provide critical insight on the rapid evolution of high redshift galaxies. This is the case of the so-called “Green Pea'' galaxies at $z\sim$0.14-0.36 \citep[GPs, ][]{cardamone_galaxy_2009}, which are low mass (M$_{*}$ < 10$^{10}$ M$_{\odot}$), highly star-forming (SFR$\sim10-60 \,M_{\odot}/yr)$) systems characterized by their unresolved and green colour appearance in Sloan Digital Sky Survey (SDSS, REF) color composite images. The GPs are characterized by their small sizes of about few kpc \citep{amorin_star_2012} and extremely compact star formation, with resolved UV effective radius ($r_{50}$) typically below 1 kpc \citep{yang_ly_2017}. Their green colour is due to their unusually strong [OIII]5007 emission, with equivalent widths EW$\sim$200-2500\AA, which is redshifted to the SDSS $r’$-band and dominates the optical emission of the galaxies. Such a strong nebular emission is characterized by its high excitation, low dust extinction and low metallicity of about 20\% solar (12$+$log(O/H)$\sim$\,8.1) in average \citep{amorin_oxygen_2010, amorin_star_2012, perez-montero_extreme_2021}. The dominant nebular emission of GPs shows high ionization conditions, which are traced by their high O32 ($\equiv[OIII]5007\AA/[OII]3727\AA$) ratios and strong emission from high-ionization lines such as $HeII\,4686\AA$ \citep{jaskot_origin_2013,ramambason_reconciling_2020}. These properties are naturally explained by galaxies experiencing a maximal ongoing starburst with young stellar ages, thus showing extremely high specific SFR (log(sSFR)$\sim$10$^{-8}$yr$^{-1}$). 

Observations of GPs with the Cosmic Origins Spectrograph (COS) mounted at the Hubble Space Telescope (HST) have allowed the detection of significant Lyman-$\alpha$ emission with escape fractions of a few to about 100\%  \citep{henry_lyupalphaemission_2015} and Lyman continuum (LyC) leakage \citep{izotov_eight_2016}, including objects showing cosmologically-relevant escape fractions that can exceed $\sim$20\% \citep[e.g.][]{izotov_j0811+4730:_2018}. Thus, these objects may provide key insight into the production and escape of ionizing photons under ISM conditions approaching those found in high-redshift systems \citep[e.g.][]{schaerer_ionizing_2016,izotov_lyman_2021}. 

More generally, the GPs are excellent laboratories to study massive star formation and feedback effects at low metallicity. \cite{amorin_star_2012} (hereafter \citetalias{amorin_star_2012}) provided a detailed analysis on three GPs: GP004054, GP113303 and GP232539 with $z\sim0.283$, $0.241$ and $0.277$. The sample was observed using the long-slit spectrograph OSIRIS at the 10.4m telescope GTC (Gran Telescopio de Canarias). This set-up represented a milestone on GPs studies, which had been mainly accomplished from the SDSS survey data. The chemical analysis, via the direct method, showed remarkable high N/O ratios, confirming  previous results by \citet{amorin_oxygen_2010} leading to the conclusion that more than 50\% of the GP sample had atypically high N/O ratios. The authors concluded that a massive inflow of metal-poor gas, as proposed in \citet{amorin_oxygen_2010}, could explain the high N/O at low O/H, and the unusually compact and strong star formation in GPs.

Similar results were later discussed by subsequent works. Some authors arrived to similar conclusions \citep{kojima_evolution_2017, loaiza-agudelo_vltx-shooter_2020} and some others argued that the high N/O may be caused, instead, by composite nebulae with different physical properties and that the presence of large amounts of Wolf-Rayet (WR) stars may favor a localized N/O enhancement due to their strong winds \citep{pilyugin_abundance_2012, hawley_abundances_2012}. Indeed, deep spectra of some high N/O GPs display prominent blue bumps suggesting a large number of WR stars within the 800 - 1200 range \citep[e.g.][]{amorin_star_2012}. 
\citet{perez-montero_are_2013} performed a spatial analysis on six WR galaxies data cubes from the PMAS (Potsdam Multi-Aperture Spectrophotometer) instrument. In four of the galaxies, the direct method results agreed with a non-uniform enhancement from WR feedback. These theoretical yields were calculated using the models from \citet{molla_modelling_2012}. The authors concluded that hydrodynamical effects (i.e., metal-poor gas inflows) are excluded as the most frequent cause of N/O excess. A comparison with model predictions led \citetalias{amorin_star_2012} to reach similar conclusions for the above three GPs. Still, later works continue to study the strong winds and possible enriched outflows from high-resolution spectra. Example of these studies on small GP samples include \citep{amorin_complex_2012,bosch_integral_2019,hogarth_chemodynamics_2020}. 

Finally, the SFH of GPs still remain largely unconstrained. It was first studied by \citetalias{amorin_star_2012}, who applied several techniques using high S/N spectra and surface brightness profiles. Two of them were based on the Simple Stellar Populations (SSPs) synthesis and one in the fitting of evolutionary synthesis models. These results led the authors to conclude that these objects were producing around $\sim4-20\%$ of their stellar mass in the current star forming burst. The ages of the evolved stellar populations resulted in a few Gyr, suggesting the presence of an old stellar host. An important achievement in this analysis was the inclusion of the nebular continuum in the analysis. In most cases, the nebular continuum can be neglected since its brightness is exceptionally low compared to the stellar one. However, in young star forming regions $(\sim10Myr)$ the gas component enhances the galaxy luminosity and tends to make the Spectral Energy Distribution (SED) redder than expected. In the \citetalias{amorin_star_2012} GPs sample, the stellar to nebular continuum percentage was between 11 to 35\% at $H\alpha$, consistent with the high $H\beta$ EWs of their star-forming regions. A recent work by \citet{clarke_old_2020} found results consistent with the above picture using HST imaging free from strong line emission. Analysing the observations with SSP models, the authors concluded that GPs have a diffuse component with colours consistent with an evolved underlying galaxy or an extended nebular continuum component.

\begin{figure*}
\includegraphics[width=1.00\textwidth]{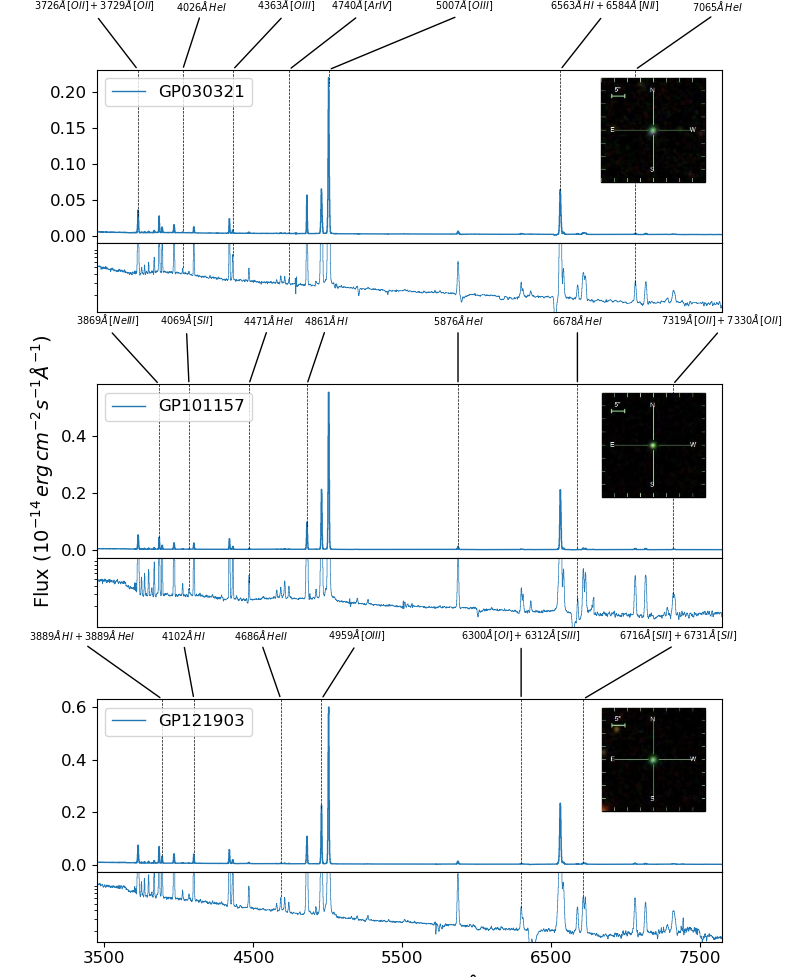}
\caption{\label{fig:sample_spectra}GP030321, GP101157 and GP121903 galaxy spectra. The lower panels are displayed in logarithmic scale, where the ordinate axis range is constrained for a better focus on the continuum features. The images on the upper-right corners belong to the SDSS sky server. Some emission lines have been labeled using the \textsc{lineidplot} library by \citet{nair_lineid_2016}}
\end{figure*}

Despite the relevance of the GPs in the bigger context of galaxy evolution and reionization and the relatively poor understanding of their own nature, high S/N optical spectra enabling detailed studies of the GPs chemical abundances and star formation histories are available. 
In the current manuscript, we build on the work by \citetalias{amorin_star_2012} by adding three new GPs to the sample. These galaxies are also observed at the GTC, with the same instrument setup but they have a lower redshift range with $\bar{z}=0.17$. Both the stellar and gas components are analysed in the current work using state-of-the-art algorithms. To begin with, both the current and \citetalias{amorin_star_2012} samples continua are treated using two techniques: In the first one, the nebular continua are calculated from first principles with a posterior SSPs synthesis in an iterative process, which considers the physical conditions measured from the galaxy emission. Secondly, the algorithm \textsc{FADO} by \citet{gomes_fitting_2017} is applied with a large stellar spectra grid. This algorithm takes into consideration the gas component making it ideal for the analysis of young star forming regions. In the gas emission analysis, the chemical composition is measured using a 14-dimensional direct method model. All these parameters are fit simultaneously using a neural networks sampler under a Bayesian inference scheme. This model is an upgrade from the algorithm presented in \citet{fernandez_bayesian_2019} for the analysis of GPs analogues in the nearby universe. Finally, using the methodology by \citet{perez-montero_photon_2020}, we compare the ionised gas emission against photoionization models to derive some physical properties of the ionising sources in these galaxies. This algorithm has been enhanced using neural networks for a better sampling of the photoionization grids. Finally, we combine both techniques in a single algorithm, which to the best author's knowledge, simultaneously explores the direct method parameter space alongside photoionization grids for the first time.

\section{Galaxy sample data} \label{data}

\begin{table*}
\caption{\label{tab:Sample-properties.} Summary log of observations}
\begin{tabu}{lcccccccc}%
\hline%
SDSS ID&Label&R.A&DEC&z&$Grisms^{1}$&$T_{exp}\,per\,grism$&Seeing&Standard stars$^{2}$\\%
\hline%
\href{http://skyserver.sdss.org/dr16/en/tools/explore/summary.aspx?plate=458\&mjd=51929\&fiber=185}{SDSSJ030321.41-075923.2}&GP030321&03:03:21.41&-07:59:23.23&0.165&R1000B, R1000R&$3\times1200s$&$0.85"$&G158-100, G158-100\\%
\href{http://skyserver.sdss.org/dr16/en/tools/explore/summary.aspx?plate=1745\&mjd=53061\&fiber=463}{SDSSJ101157.08+130822.0}&GP101157&10:11:57.08&+13:08:22.06&0.144&R1000B, R1000R&$3\times1200s$&$1"$&Ross640, G191-B2B\\%
\href{http://skyserver.sdss.org/dr16/en/tools/explore/summary.aspx?plate=1766\&mjd=53468\&fiber=557}{SDSSJ121903.98+152608.5}&GP121903&12:19:03.98&+15:26:08.53&0.196&R1000B, R1000R&$3\times1200s$&$0.9"$&Feige 34, Ross640\\%
\hline%
\end{tabu}

\raggedright{\footnotesize{$^1$ The science targets were observed with a 0.8"$\times$7.4' slit width. $^2$ The standard stars were observed with a 2.5"$\times$7.4' slit width.}}\\
\end{table*}

The sample consists in three galaxies selected among the brightest GPs in the sample presented by \cite{cardamone_galaxy_2009}. Their SDSS references are: 
\href{http://skyserver.sdss.org/dr16/en/tools/explore/summary.aspx?plate=458&mjd=51929&fiber=185}{SDSSJ030321.41-075923.2},  \href{http://skyserver.sdss.org/dr16/en/tools/explore/summary.aspx?plate=1745&mjd=53061&fiber=463}{SDSSJ101157.08+130822.0} and \href{http://skyserver.sdss.org/dr16/en/tools/explore/summary.aspx?plate=1766&mjd=53468&fiber=557}{SDSSJ121903.98+152608.5}. The first GP can also be found in the SHOC catalogue (SDSS HII-galaxies with Oxygen abundances Catalog) by \citet{kniazev_strong_2004} as SHOC148. For simplicity's sake, however, the GPs will be referred within this manuscript as GP030321, GP101157 and GP121903. 

Observations were conducted in service mode at the Gran Telescopio de Canarias (GTC) in Roque de los muchachos, La Palma, Spain. with the  Optical System for Imaging and low-Intermediate-Resolution Integrated Spectroscopy (OSIRIS) under program GTC63-10B (PI: Amor\'in) with telescope time granted in  09/2010, 10/2010, 12/2010. We used the R1000B and R1000R grisms with a  resolving power R$\approx$1018  at $\lambda_{c,\,Blue}\approx5510\AA$ and R$\approx$R1122 at $\lambda_{c,\,Red}\approx7510\AA$. The wavelength ranges for these grisms are $3300-7500\AA$ and $5100-10250\AA$, respectively. In the case of GP030321, the red arm displayed contamination from the the second order light from the blue side. Due to this and severe telluric contamination, we decided to crop the spectra redder than $\lambda > 7600\AA$. We choose a slit width of 0.8" resulting in a full width at half maximum (FWHM) around 3 pixels. The observations were conducted under average seeing conditions of 1.00". Three exposures of $3\times1200s$ per grating and object, were obtained. Standard calibrations (bias, dome flats) exposures and HgAr+Xe+Ne lamp arcs were obtained after each science exposure. Two spectrophotometric stars and arcs for flux calibration were also  obtained. A summary log of observations can be found in Table  \ref{tab:Sample-properties.}

We use \textsc{IRAF}\footnote{\textsc{iraf}: The image Reduction and Analysis Facility is distributed by the National Optical Astronomy Observatory, which is operated by the Association of Universities for Research in astronomy (AURA), Inc., under the cooperative agreement with the National Science Foundation (\href{http://iraf.noao.edu/}{http://iraf.noao.edu/})} to reduce the OSIRIS spectra following \citetalias{amorin_star_2012}. In brief, after standard calibrations (bias+overscan subtraction, flat fielding and cosmic-ray removal) we perform the wavelength and flux calibration. The final spectra wavelength array has an accuracy of $0.1\AA$ measured from bright sky lines. Flux calibration, instead, has large uncertainty at $\lambda < 3500\AA$ due to low signal at the CCD edge. This blue edge of the spectra was not considered in the subsequent analysis. Finally, the flux calibration resulted in an uniform offset between the blue and red arms. In the three galaxies, the overlapping continuum flux sigma is of the same order. This was corrected by applying a constant offset obtained after scaling the two arms using the overlapping regions between $\sim 5100\AA$ and $6000\AA$. Additionally, emission lines were normalized using a recombination flux within their arm wavelength range (see section \ref{sec:extinction_calc}).

\begin{table*}
\caption{\label{tab:emission_fluxes_gp03032} Emission line fluxes for the GPs sample. Column (1) Wavelengths in $\AA$ at rest frame. The labels with a "g" subscript identify emission lines whose measurement consisted in a Gaussian profile deblending. Column (2) Reddening curve. Columns (3-4), (5-6), (7-8) display the observed ($F(\lambda)$) and extinction corrected fluxes relative to $F(H\beta) = I(H\beta) = 1000$. At the bottoms of the table the logarithmic extinction coefficient, the equivalent width and the observed flux for $H\beta$ are provided}
\centering{\begin{tabu}{lccccccc}%
\hline%
&&\multicolumn{2}{c}{GP030321}&\multicolumn{2}{c}{GP101157}&\multicolumn{2}{c}{GP121903}\\%
Line label&$f_{\lambda}$&$F(\lambda)$&$I(\lambda)$&$F(\lambda)$&$I(\lambda)$&$F(\lambda)$&$I(\lambda)$\\%
\hline%
$3704\AA\,HI$&0.33&-&-&$5\,\pm\,1$&$10\,\pm\,2$&$8\,\pm\,1$&$9\,\pm\,1$\\%
$3712\AA\,HI$&0.33&-&-&$5\,\pm\,1$&$9\,\pm\,2$&$5\,\pm\,1$&$6\,\pm\,1$\\%
$3726\AA\,[OII]+3729\AA\,[OII]$&0.32&$654\,\pm\,2$&$978\,\pm\,94$&$487\,\pm\,1$&$928\,\pm\,150$&$607\,\pm\,3$&$667\,\pm\,11$\\%
$3750\AA\,HI$&0.32&$14\,\pm\,1$&$21\,\pm\,3$&$19\,\pm\,2$&$36\,\pm\,6$&$26\,\pm\,1$&$29\,\pm\,2$\\%
$3771\AA\,HI$&0.31&$17\,\pm\,2$&$25\,\pm\,3$&$26\,\pm\,1$&$48\,\pm\,8$&$40\,\pm\,1$&$43\,\pm\,1$\\%
$3798\AA\,HI$&0.31&$36\,\pm\,1$&$53\,\pm\,5$&$32\,\pm\,2$&$59\,\pm\,10$&$57\,\pm\,2$&$63\,\pm\,2$\\%
$3820\AA\,HeI$&0.30&$6\,\pm\,1$&$9\,\pm\,2$&$8\,\pm\,1$&$15\,\pm\,3$&$5\,\pm\,2$&$5\,\pm\,2$\\%
$3835\AA\,HI$&0.30&$69\,\pm\,2$&$100\,\pm\,9$&$44\,\pm\,1$&$80\,\pm\,12$&$79\,\pm\,1$&$86\,\pm\,2$\\%
$3869\AA\,[NeIII]$&0.29&$359\,\pm\,6$&$517\,\pm\,45$&$348\,\pm\,1$&$624\,\pm\,91$&$458\,\pm\,6$&$498\,\pm\,9$\\%
$3889\AA\,HI+3889\AA\,HeI$&0.29&$159\,\pm\,2$&$227\,\pm\,20$&$117\,\pm\,1$&$207\,\pm\,30$&$200\,\pm\,2$&$217\,\pm\,4$\\%
$3970\AA\,HI$&0.27&$260\,\pm\,1$&$363\,\pm\,29$&$220\,\pm\,1$&$374\,\pm\,50$&$326\,\pm\,1$&$353\,\pm\,5$\\%
$4026\AA\,HeI$&0.25&$16\,\pm\,1$&$22\,\pm\,2$&$12\,\pm\,0.5$&$19\,\pm\,3$&$19\,\pm\,1$&$21\,\pm\,1$\\%
$4069\AA\,[SII]$&0.24&$9\,\pm\,1$&$12\,\pm\,1$&$7\,\pm\,0.5$&$12\,\pm\,2$&$13\,\pm\,0.5$&$13\,\pm\,1$\\%
$4102\AA\,HI$&0.23&$185\,\pm\,2$&$247\,\pm\,17$&$179\,\pm\,1$&$283\,\pm\,33$&$277\,\pm\,1$&$296\,\pm\,4$\\%
$4341\AA\,HI$&0.16&$403\,\pm\,2$&$490\,\pm\,23$&$320\,\pm\,1$&$438\,\pm\,34$&$459\,\pm\,1$&$480\,\pm\,4$\\%
$4363\AA\,[OIII]$&0.15&$89\,\pm\,1$&$107\,\pm\,5$&$86\,\pm\,1$&$116\,\pm\,9$&$132\,\pm\,1$&$138\,\pm\,2$\\%
$4471\AA\,HeI$&0.12&$33\,\pm\,1$&$38\,\pm\,2$&$24\,\pm\,0.5$&$30\,\pm\,2$&$44\,\pm\,0.5$&$45\,\pm\,1$\\%
$4658\AA\,[FeIII]$&0.06&$7\,\pm\,1$&$7\,\pm\,1$&$10\,\pm\,0.5$&$11\,\pm\,0.5$&$11\,\pm\,0.5$&$11\,\pm\,0.5$\\%
$4686\AA\,HeII$&0.05&$20\,\pm\,1$&$22\,\pm\,1$&$14\,\pm\,0.5$&$16\,\pm\,1$&$25\,\pm\,0.5$&$26\,\pm\,1$\\%
$4711\AA\,[ArIV]+4713\AA\,HeI$&0.04&$17\,\pm\,1$&$18\,\pm\,1$&$24\,\pm\,0.5$&$26\,\pm\,1$&$22\,\pm\,1$&$23\,\pm\,1$\\%
$4740\AA\,[ArIV]$&0.03&$7\,\pm\,1$&$7\,\pm\,1$&$13\,\pm\,0.5$&$14\,\pm\,0.5$&$14\,\pm\,1$&$14\,\pm\,1$\\%
$4861\AA\,H\beta$&0.00&$1000\,\pm\,2$&$1000\,\pm\,2$&$1000\,\pm\,1$&$1000\,\pm\,1$&$1000\,\pm\,2$&$1000\,\pm\,2$\\%
$4922\AA\,HeI$&-0.02&$10\,\pm\,0.5$&$10\,\pm\,0.5$&$9\,\pm\,0.5$&$9\,\pm\,0.5$&$7\,\pm\,1$&$7\,\pm\,1$\\%
$4959\AA\,[OIII]$&-0.03&$1742\,\pm\,2$&$1687\,\pm\,13$&$2264\,\pm\,2$&$2150\,\pm\,28$&$2124\,\pm\,4$&$2108\,\pm\,5$\\%
$4987\AA\,[FeIII]$&-0.03&-&-&$6\,\pm\,2$&$6\,\pm\,2$&-&-\\%
$5007\AA\,[OIII]$&-0.04&$5801\,\pm\,7$&$5533\,\pm\,62$&$6553\,\pm\,4$&$6075\,\pm\,115$&$6094\,\pm\,7$&$6027\,\pm\,13$\\%
$5048\AA\,HeI$&-0.05&$5\,\pm\,2$&$4\,\pm\,2$&$1\,\pm\,0.5$&$1\,\pm\,0.5$&$3\,\pm\,0.5$&$3\,\pm\,0.5$\\%
$5200\AA\,[NI]$&-0.08&$3\,\pm\,1$&$3\,\pm\,1$&$5\,\pm\,0.5$&$4\,\pm\,0.5$&$3\,\pm\,0.5$&$3\,\pm\,0.5$\\%
$5270\AA\,[FeIII]$&0.28&-&-&-&-&$5\,\pm\,0.5$&$6\,\pm\,0.5$\\%
$5272\AA\,[FeIII]$&0.28&-&-&$4\,\pm\,0.5$&$8\,\pm\,1$&-&-\\%
$5518\AA\,[ClIII]$&0.22&-&-&$2\,\pm\,0.5$&$4\,\pm\,1$&$3\,\pm\,0.5$&$3\,\pm\,0.5$\\%
$5538\AA\,[ClIII]$&0.21&-&-&$2\,\pm\,0.5$&$3\,\pm\,0.5$&$1\,\pm\,0.5$&$1\,\pm\,0.5$\\%
$5755\AA\,[NII]$&0.16&$4\,\pm\,1$&$4\,\pm\,1$&-&-&-&-\\%
$5876\AA\,HeI$&0.13&$191\,\pm\,2$&$226\,\pm\,9$&$124\,\pm\,1$&$163\,\pm\,11$&$121\,\pm\,1$&$126\,\pm\,1$\\%
$6300\AA\,[OI]_g$&0.05&$46\,\pm\,2$&$49\,\pm\,2$&$24\,\pm\,0.5$&$27\,\pm\,1$&$28\,\pm\,0.5$&$29\,\pm\,0.5$\\%
$6312\AA\,[SIII]_g$&0.05&$22\,\pm\,2$&$24\,\pm\,2$&$15\,\pm\,0.5$&$16\,\pm\,1$&$11\,\pm\,0.5$&$11\,\pm\,0.5$\\%
$6364\AA\,[OI]$&0.04&$19\,\pm\,1$&$20\,\pm\,1$&$8\,\pm\,0.5$&$9\,\pm\,0.5$&-&-\\%
$6548\AA\,[NII]_g$&0.00&-&-&$24\,\pm\,8$&$24\,\pm\,8$&$47\,\pm\,5$&$47\,\pm\,5$\\%
$6563\AA\,H\alpha_g$&0.00&$2791\,\pm\,26$&$2791\,\pm\,26$&$2795\,\pm\,27$&$2795\,\pm\,27$&$2777\,\pm\,19$&$2777\,\pm\,19$\\%
$6584\AA\,[NII]_g$&-0.00&$148\,\pm\,20$&$147\,\pm\,20$&$72\,\pm\,19$&$71\,\pm\,19$&$139\,\pm\,13$&$138\,\pm\,13$\\%
$6678\AA\,HeI$&-0.02&$51\,\pm\,1$&$50\,\pm\,1$&$16\,\pm\,0.5$&$16\,\pm\,0.5$&$34\,\pm\,1$&$34\,\pm\,0.5$\\%
$6716\AA\,[SII]_g$&-0.03&$168\,\pm\,3$&$162\,\pm\,3$&$73\,\pm\,1$&$69\,\pm\,1$&$54\,\pm\,1$&$54\,\pm\,1$\\%
$6731\AA\,[SII]_g$&-0.03&$95\,\pm\,2$&$91\,\pm\,3$&$61\,\pm\,1$&$57\,\pm\,1$&$52\,\pm\,1$&$52\,\pm\,1$\\%
$7065\AA\,HeI$&-0.09&$78\,\pm\,1$&$70\,\pm\,2$&$62\,\pm\,1$&$51\,\pm\,3$&$62\,\pm\,1$&$60\,\pm\,1$\\%
$7136\AA\,[ArIII]$&-0.11&$87\,\pm\,1$&$76\,\pm\,3$&$66\,\pm\,1$&$53\,\pm\,3$&$47\,\pm\,0.5$&$46\,\pm\,1$\\%
$7281\AA\,HeI$&-0.14&$10\,\pm\,1$&$8\,\pm\,1$&$6\,\pm\,0.5$&$4\,\pm\,0.5$&$6\,\pm\,1$&$6\,\pm\,1$\\%
$7319\AA\,[OII]+7330\AA\,[OII]$&-0.14&$66\,\pm\,1$&$55\,\pm\,2$&$41\,\pm\,1$&$31\,\pm\,2$&$46\,\pm\,1$&$44\,\pm\,1$\\%
$7751\AA\,[ArIII]$&-0.22&$28\,\pm\,0.5$&$21\,\pm\,1$&$25\,\pm\,2$&$16\,\pm\,2$&$11\,\pm\,0.5$&$10\,\pm\,0.5$\\%
\hline%
$c(H\beta)$&&\multicolumn{2}{c}{$0.54\,\pm\,0.13$}&\multicolumn{2}{c}{$0.87\,\pm\,0.22$}&\multicolumn{2}{c}{$0.13\,\pm\,0.02$}\\%
$-W(\beta)(\AA)$&&\multicolumn{2}{c}{$115$}&\multicolumn{2}{c}{$274$}&\multicolumn{2}{c}{$217$}\\%
$F(H\beta) (10^{14} \cdot erg\,cm^{-2} s^{-1})$&&\multicolumn{2}{c}{$0.383$}&\multicolumn{2}{c}{$0.826$}&\multicolumn{2}{c}{$0.945$}\\%
\hline%
\end{tabu}}
\end{table*}

\section{Analysis of the spectra} \label{treatment}

In Fig. \ref{fig:sample_spectra}, we present the OSIRIS spectra of the three observed galaxies in the rest frame. The lower panels show the spectrum in logarithmic scale centred on the continuum region. This helps to appreciate the weak emissions,
such as the $[OIII]4363\AA$ and $[SIII]6312\AA$ auroral lines and the $HeII4686\AA$ transition, which are prominent in the three galaxies. Moreover, this scale also makes it easier to identify the Balmer jump with its higher continuum flux below the $3646\AA$ mark. The plot insets show the colour SDSS cutout of the galaxies from the SDSS sky server. Both the spectra and the images illustrate the characteristic properties of GP galaxies, namely unresolved morphology in SDSS and green colour due to extreme nebular emission, in particular high $EW([OIII]4959,5007\AA)$ and emission lines falling in the r'-band.
In this section, we describe the procedure employed to disentangle the main phenomena contributing to the observed radiation. These are the light extinction from the dust particles, the nebular continua from the hydrogen and helium atoms, the stellar continua from the underlying galaxy and the emission lines from the ionised gas. Moreover, this section also provides a brief technical explanation of the applied methodology, similar to the approach used in \citet{fernandez_determination_2018} for the high precision measurement of the primordial helium abundance using HII galaxies.  It should be explained that, in order to guarantee consistent results, some of the steps described below were repeated iteratively, as the gas extinction and the chemical composition are used as inputs in the nebular and stellar continua fitting. This treatment assumes that gas and stars have the same light attenuation behind a screen of dust \citep[see][]{calzetti_dust_2001}. A more realistic scenario assumes clumps of dust with uneven extinction. Ignoring the disparity between the stars and gas attenuation can lead to an overestimation of the stellar and nebular continua on the line of sight. Fortunately, these objects have very low logarithmic extinction coefficients, making this simple model suitable.

\subsection{Emission-line flux measurement}

The spectra analysis starts with the measurement of the line fluxes. For every line, three spectral regions are selected: One covering the line width and two the adjacent continua. For a single emission line (for example $[OIII]5007\AA$) or an emission line consisting in several ionic transitions, which cannot be deblended (for example $[OII]3726\AA,3720\AA$ at our resolution) the measurement is done via a Monte Carlo integration: In a 1000 iterations loop, each pixel in the emission line is added a random quantity from a normal distribution. This distribution is centred at zero and its sigma is computed from the standard deviation of the adjacent continua regions (assuming a linear relation between the flux and the wavelength). Afterwards, we integrate the pixels within the emission line region. The mean value from the 1000 iterations is taken as the line flux measurement, while its standard deviation quantifies the flux uncertainty. 
Emission lines consisting in more than one ionic transitions and whose components can be deblended, are fitted with a Gaussian mixture using \textsc{LMfit} by \citet{newville_lmfit_2014}. In the sample spectra, the blended lines are $[OI]6300\AA-[SII]6312\AA$, $H\alpha-[NII]6548\AA,6584\AA$ and $[SII]6716\AA,6731\AA$. The theoretical Gaussian area $\left(A \cdot \sqrt{2 \pi}\cdot\sigma\right)$ quantifies the emission flux, where $A$ is the Gaussian curve amplitude (the peak height with respect to the line continuum) and $\sigma$ is the Gaussian curve standard deviation. The uncertainty in the measurement is taken from \textsc{LMfit} $1\sigma$ standard error output from the default Levenberg-Marquart algorithm. We, however, are not using the default Gaussian parametrisation from \textsc{LMfit}. Instead, we define our own model which fits $A_{i}$, $\mu_{i}$ (the peak central wavelength) and  $\sigma_{i}$ for every Gaussian component $(i)$.

In the first iteration, the emission flux measurement is performed on the observed spectra, Fig. \ref{fig:sample_spectra}. On subsequent iterations, once the stellar continuum has been fitted, this component is removed from the observation before the flux measurement. This provides a correction for the stellar absorption on the recombination lines. Table \ref{tab:emission_fluxes_gp03032} presents the emission line fluxes obtained from the output of the final iteration consisting on the pure emission. At the bottom of the table, the logarithmic extinction coefficient $c(H\beta)$ along with the $H\beta$ line equivalent width and flux is tabulated for each object. 

\subsection{Interstellar reddening correction}\label{sec:extinction_calc}

The radiation reaching us is affected by dust particles, which scatter the light of the star forming region. A correction for this extinguished light can be accomplished by comparing the observed relative fluxes from recombination lines against their corresponding emissivity ratios. Any significant deviation can be attributed to dust scattering. This is a safe assumption since these lines have a very weak dependence on the electron temperature and density. This condition also assumes the absorption from the underlying stellar population is negligible. The extinction law selected for this analysis was \cite{cardelli_relationship_1989} with $R_V = 3.1$. In general, the extinction law choice has a weak impact on the optical spectrum analysis of GPs. This is due to their relative low dust content \citep[see][]{cardamone_galaxy_2009, amorin_oxygen_2010}. This reddening law is also the one applied on the SSP synthesis (see section \ref{sec:stellar-calc}).

As it was mentioned in Sec. \ref{data}, our observations display a flux mismatch between the blue and red grisms. To avoid inconsistencies in measurements from both grisms, the following scheme was applied: The emission lines at each range are normalised by the brightest hydrogen line ($H\beta$ for the blue arm and $H\alpha$ for the red arm). The logarithmic extinction coefficient, $c(H\beta)$, is calculated with the three most intense hydrogen lines in the blue arm ($H\beta$, $H\gamma$ and $H\delta$). Finally, in order to apply the extinction correction in the red arm features the classical formulation was adjusted to:
\begin{equation}
\label{eq:reddening_cor}
\frac{I_{\lambda}}{I_{H\beta}}=\frac{F_{\lambda}}{F_{H\alpha}}\cdot\frac{\epsilon_{H\alpha}}{\epsilon_{H\beta}}10^{c\left(H\beta\right)\left(f_{\lambda}-f_{H\alpha}\right)}
\end{equation}
where $F(H\lambda)$, $I(H\lambda)$ are the line emission flux and intensity respectively, $\frac{\epsilon_{H\alpha}}{\epsilon_{H\beta}}$ is the theoretical ratio between the $H\alpha$ and $H\beta$ emissivities, $c(H\beta)$ is the logarithmic extinction coefficient and $f(\lambda)$ is the transition reddening law value at the corresponding emission wavelength. The $f(\lambda)$ coefficient for each line can be found in Table \ref{tab:emission_fluxes_gp03032}. The emissivities were calculated using the atomic data references displayed in Table \ref{tab:atomic-data} from the \textsc{PyNeb} library by \citet{luridiana_pyneb:_2015}. In the first iteration, the recombination emissivity ratios were calculated using standard conditions of $T_e=10000K$ and $n_e=100\,cm^{-3}$. In subsequent iterations, the density and the low ionisation region temperature were used instead. 
The previous two-arm reddening correction was also applied in the continuum analysis. In this case, however, Eq. \ref{eq:reddening_cor} was adapted for an absolute flux correction instead of a relative one. 

\subsection{Measurement and treatment of the nebular continuum} \label{sec:nebular_calc}

In \citetalias{amorin_star_2012} the nebular continuum was characterised via the \textsc{PopStar} evolutionary synthesis models \cite[see ][]{molla_popstar_2009}. These grids define star forming regions consistently with both stellar atmospheres and ionised nebular continua. In the current analysis, we use the same atomic data and similar methodology as in \cite{molla_popstar_2009}. The main difference is that the current analysis computes the nebular continuum from the measured gas properties, instead of using the temperature assigned to the star forming region according to the metallicity \citep[see Table 6 in][]{molla_popstar_2009}. The full procedure description can be found in \cite{fernandez_determination_2018} but the main steps are summarised as follows:

 \begin{itemize}
   \item  The continuous emission coefficient, $\gamma_\nu$ $\left(erg\,\nicefrac{cm^3}{s}\right)$, is calculated for the Free-Free (FF), Free-Bound (FB) and Bound-Bound (BB) continua. The Bremsstrahlung or FF continuum computation sticks to the classical methodology described by \cite{brown_theoretical_1970} and \cite{osterbrock_astrophysics_1974} for the $H^{+}$, $He^{+}$ and $He^{2+}$ ions. The FB continuum is calculated from the tabulated data by \cite{ercolano_theoretical_2006} again for the hydrogen and helium atoms. Finally, the BB or two-photon continuum is calculated via the analytical expression by \cite{nussbaumer_hydrogenic_1984} for the hydrogen $2s^{2}S_{1/2}\rightarrow1s^{2}S_{1/2}$ decay. In the first iteration, these components are calculated using standard conditions: $T_e=10000K$, $n_e=100\,cm^-3$, $y^+=0.01$ and $y^{2+}=0.001$. In the following iteration, the physical conditions are those measured from each GP low ionization region.
   \item  The combined continuous emission spectrum is flux calibrated using a \cite{zanstra_untersuchungen_1931} like approach using the $H\alpha$ flux and the effective recombination coefficient from \cite{pequignot_total_1991}, $\alpha^{eff}_{\alpha}$.
   \item The output spectrum is artificially reddened using the logarithmic extinction coefficient previously measured. Finally, this continuum is removed from the observed spectrum.
 \end{itemize}

\subsection{Measurement and treatment of the stellar continuum}\label{sec:stellar-calc}

Two different approaches were considered in the analysis of the GPs continua. In the first one the nebular continuum is removed from the observed spectrum prior to a SSP synthesis using \textsc{Starlight} by \cite{fernandes_semi-empirical_2005}. This step is essential for an unbiased study of the physical and evolutionary properties 
of the stellar component of starburst galaxies, given that the nebular continuum shows, contrary to the young ionizing stellar component, a nearly flat SED, thus becoming progressively important in the red spectral range \citep[see][]{krueger_optical_1995,papaderos_age_1998,izotov_green_2011}. Adhering to STARLIGHT manual recommendations, the input spectrum is resampled to $1\AA$ per pixel. Additionally, an input mask file is provided to exclude the emission lines, sky contamination and noisy regions from the fitting. The input stellar bases combine grids from \cite{bruzual_stellar_2003}, \cite{falcon-barroso_updated_2011} and \cite{delgado_evolutionary_2005}. The 293 spectra grid covers a $\nicefrac{Z_{\odot}}{200}$ - $1.5Z_{\odot}$ stellar metallicity range and a $1 Myr$ - $17 Gyr$ stellar age range. The spectral range in these bases goes from $3300\AA$ to $6990\AA$ providing a good match to our observational wavelength range. The \textsc{Starlight} example configuration file was used as a template for our measurements with the following modifications: The normalization window chosen was $4760\AA$ to $4835\AA$. The minimum visual extinction was set to zero, while the maximum stellar velocity dispersion was defined equal to the gas velocity dispersion measured from the $[OIII]5007\AA$ line. Finally, the stellar velocity in the line of sight was fixed during the fits. This was done to avoid issues with the stellar velocity in the line of sight: since these spectra have no noticeable absorptions, enabling this feature results in a mismatch between the predicted absorption features and the observed emission lines.

The second methodology relies on the population spectral synthesis code \textsc{FADO} (Fitting Analysis Using Differential evolution Optimization) by \citet{gomes_fitting_2017}. A unique feature of this algorithm is that it takes both stellar and nebular emission into account and, quite importantly, identifies the SFH that self-conistently reproduces the observed nebular characteristics of a star-forming galaxy (H$\alpha$ and H$\beta$  luminosities and equivalent widths, shape of the continuum around the Balmer and Paschen jump). Another novel feature of FADO is the use of genetic algorithms for multi-objective optimization. The hydrogen and helium atomic data sources cited in section \ref{sec:nebular_calc} are also used in \textsc{FADO}. Moreover, its calculation follows a similar approach: The electron temperature and density are computed from the $[OIII]$ and $[SII]$ transitions while the nebular extinction is computed from the $H\alpha$ and $H\beta$ fluxes. Afterwards, the SSPs synthesis is applied considering the nebular component. In this fitting, the SSP library considered is the reviewed grid by \cite{bruzual_stellar_2003}. It consists in 1098 bases with a wider time range at smaller increments. In the $5 < log(age) < 6$ interval there are 18 steps with $\Delta(log(age)=0.05$. For $6 < log(age) < 10.3$, there are 202 steps at $\Delta(log(age))=0.02$. The stellar metallicities included are $Z = 0.005$, $0.020$, $0.200$, $0.400$, $1.000$ and $2.500$ in a $Z_{\odot}$ units. 

In the continua analysis, we incorporate the GP sample from \citetalias{amorin_star_2012}. This is done to test the previous results against those using the novel stellar libraries and fitting algorithm.

\subsection{Physical properties and chemical abundances analysis} \label{chemical_analysis}

\begin{table}
\caption{\label{tab:Priors-and-likelihood} Priors and likelihood distributions
in our model. The term $X^{i+}$ includes all the ionic metal abundances:
$Ar^{2+}$, $Ar^{3+}$, $Cl^{3+}$, $Fe^{3+}$, $O^{+}$, $O^{2+}$, $Ne^{3+}$,
$N^{+}$, $S^{+}$, $S^{2+}$, $y^{+}$ and $y^{2+}$. The helium abundances are defined in logarithmic scale, while the metals are defined using a $12+log\left(X^{i+}\right)$ notation.}
\centering{\begin{tabular}{cc}
\toprule 
Parameter & Prior distribution\tabularnewline
\midrule
$T_{low}$ & $Normal(\mu=15000\,K,\,\sigma=5000\,K)$\tabularnewline
$T_{high}$ & $Normal(\mu=15000\,K,\,\sigma=5000\,K)$\tabularnewline
$n_{e}$ & $HalfCauchy\left(\mu=2.0,\,\sigma=0\right)$\tabularnewline
$c(H\beta)$ & $HalfCauchy\left(\mu=2.0,\,\sigma=0\right)$\tabularnewline
$X^{i+}$ & $Normal(\mu=5,\,\sigma=5)$\tabularnewline
$y^{+}$ & $Normal\left(\mu=0,\,\sigma=3\right)$\tabularnewline
$y^{2+}$ & $Normal\left(\mu=0,\,\sigma=3\right)$\tabularnewline
$T_{eff}$ & $Uniform\left(min=30000K,\,max=90000K\right)$\tabularnewline
$log(U)$ & $Uniform\left(min=-4.0,\,max=-1.0\right)$\tabularnewline
\midrule 
Parameter & Likelihood distribution\tabularnewline
\midrule
$\frac{F_{X^{i+},\,\lambda}}{F_{H\beta}}$ & $Normal(\mu=\frac{F_{X^{i+},\,\lambda,\,obs}}{F_{H\beta}},\sigma=\frac{\sigma_{X^{i+},\,\lambda,\,obs}}{F_{H\beta}})$\tabularnewline
\bottomrule
\end{tabular}
}
\end{table}

The gas chemical abundance analysis using optical emission lines has been recently illustrated by \cite{perez-montero_ionized_2017}. In the scenario where one or several electron temperatures and densities can be derived, this methodology is known as the direct method. This is the ideal case, since it makes possible to calculate each transition emissivity. However, since the electron temperature measurement depends on weak auroral lines, a high precision result requires a high S/N. A simple model predicts the relation between the phenomena contributing to the recombination and collisionally excited line flux: 

\begin{equation}
\frac{F_{X^{i+},\,\lambda}}{F_{H\beta}}=X^{i+}\frac{\epsilon_{X^{i+},\,\lambda}\left(T_{e},\,n_{e}\right)}{\epsilon_{H\beta}\left(T_{e},\,n_{e}\right)}\cdot10^{-c\left(H\beta\right)\cdot f_{\lambda}}\label{eq:fluxFormula}
\end{equation}
where $\nicefrac{\epsilon_{X^{i+},\,\lambda}}{\epsilon_{H\beta}}$ is the relative emissivity at the transition wavelength $\lambda$, for an ion with abundance $X^{i+}$, at certain electron temperature $T_{e}\,\left(K\right)$ and electron density $n_{e}$ $\left(cm^{-3}\right)$. The term $c\left(H\beta\right)$ is the relative logarithmic extinction coefficient at $H\beta$ for a reddening law $f_{\lambda}$.

For this study, we solve simultaneously this system of equations via a Bayesian model which was presented  in \cite{fernandez_bayesian_2019}. This algorithm was written using the probabilistic programming library \textsc{PyMC3} by \citet{salvatier_probabilistic_2016}. Its NUTs (No-U-Turns) sampler \citep[see][]{hoffman_no-u-turn_2011} is based on a HMC (Hamiltonian Monte-Carlo) algorithm, written itself via the neural networks package Theano \citep[see][]{the_theano_development_team_theano:_2016}. The first advantage from this approach is the complete fitting of the parameter space. In the current work, this means 14 out of the current maximum of 16 parameters: One electron temperature, one electron density, the logarithmic extinction coefficient and eleven ionic species, see Table \ref{tab:Priors-and-likelihood}. The second advantage is its measurement speed: Fitting up to 23 emission line fluxes from the input GPs spectra takes less than 160 seconds in a i7-10700K Processor. As it was discussed in \cite{fernandez_bayesian_2019} the results are consistent with those from the traditional method. The full algorithm description, as well as examples showing its convergence accuracy and stability, can be found in \cite{fernandez_bayesian_2019}. However, the following points provide a recap on its workings as well as an explanation on the updates:
\begin{equation}
Pr\left(\theta|y\right)=\frac{Pr\left(y|\theta\right)Pr\left(\theta\right)}{Pr\left(y\right)}\label{eq:bayesTheorem}
\end{equation}

\begin{itemize}
    \item The first step in a Bayesian model involves the priors selection. These distributions, $Pr\left(\theta\right)$ in Eq. \ref{eq:bayesTheorem}, represent our knowledge of the physical parameters prior to their measurement. Ideally, the user introduces an "uninformative" prior: This means a wide distribution, which guarantees a large range of solutions for the parameter value. As the sampling process reaches the solution region, the value of $Pr\left(\theta\right)$ changes very little and the simulation becomes dominated by the likelihood, $Pr\left(y|\theta\right)$. Table \ref{tab:Priors-and-likelihood} presents the configuration for our priors. The main novelty with respect to those chosen in \cite{fernandez_bayesian_2019} is the inclusion of priors for the  $Cl^{3+}$,  $Ne^{3+}$ and $Fe^{3+}$ abundances. These ionic abundances measurement is possible via the $[ClIII]5518\AA$, $[NeIII]3968\AA$ and $[FeIII]4658\AA$ lines respectively. Moreover, both the electron density, ne, and the reddening constant, $c(H\beta)$ use a Half-Cauchy prior distribution to avoid non-physical negative values for these parameters.
    \item The likelihood, $Pr\left(y|\theta\right)$ in Eq. \ref{eq:bayesTheorem}, evaluates the observational data against the sampled parameter values at each simulation step. As it was discussed in \cite{fernandez_bayesian_2019}, the Bayesian paradigm cannot interpret uncertainty on the input data. In this astrophysical model, however, this is critical. For example, while the uncertainty in the photon flux from the $[OIII]5007\AA$ transition may be neglected, the same cannot be said for photons flux measured in the $[OIII]4363\AA$ transition. Moreover, not all the data points have the same impact on the parameter space exploration. Indeed, the $[OIII]5007\AA$ photons mainly contribute to the $O^{2+}$ abundance measurement. In contrast, the $[OIII]4363\AA$ photons have a strong impact in the temperature measurement, and therefore, all the metals abundance. To account for the uncertainty weight, a normal distribution likelihood is defined for each input emission line. This distribution is centred at the current iteration synthetic flux, $F_{\chi^{i+},\lambda}$, while its width is given by the observed relative flux uncertainty, $\sigma_{\chi^{i+},\lambda}$. In this design, emission lines with a larger uncertainty display a wider likelihood, which translates into a larger acceptance rate.

    \item This chemical model considers two ionization regions: A high ionization region for the $Ar^{3+}$, $Cl^{3+}$, $Fe^{3+}$, $O^{2+}$, $y^{+}$ and $y^{2+}$ ions the low ionization region for the remaining ions. The main temperature diagnostic ratio in the sample wavelength range is $R_{[OIII]} = \nicefrac{I(4959\AA)+I(5007\AA)}{I(4363\AA)}$. This is the temperature used for the high ionization region, $T_{high} = T_{[OIII]}$. For the low ionization region, we use the $O^2+$ temperature, $T_{low} = T_{[OII]}$.  However, since the  $[OII]7319,7330\AA$ have a poor S/N in our spectra we use the relation presented in \cite{perez-montero_impact_2009}:
    \begin{equation}
    T[OII]=\frac{1.2+0.002n_{e}+\frac{4.2}{n_{e}}}{\frac{10000}{T[OIII]}+0.08+0.003n_{e}+\frac{2.5}{n_{e}}} \label{eq:TOIII-TSIII-relation}
    \end{equation} 
    where $n_{e}$ is in $cm^{-3}K$ units. This equation was computed from the models in \citet{perez-montero_line_2003} using \textsc{Cloudy v96} \citep[see][]{ferland_hazy_2002}. These models agree with the observations, in a non-linear relation between both oxygen ions temperature. The models assume that the electron density remains spatially constant. This may not be realistic for SFR with complex dynamics. Additionally, it should be remembered that the electron density below $100cm^{-3}$ cannot be accurately measured with the available diagnostics. This means that keeping $n_{e}=100cm^{-3}$ constant at this boundary may be underestimating the predicted low ionization temperature. In this work, we use the ratio $R_{[SII]} = \nicefrac{I(6716\AA)}{I(6731\AA)}$ to anchor the electron density. 
    \item In \cite{fernandez_bayesian_2019}, it was discussed that the main disadvantage from the Bayesian sampler was the requirement to parameterize the emissivity temperature-density plane. In contrast to the traditional Monte Carlo algorithm used in  \cite{fernandez_determination_2018}, tensors in neural networks cannot be easily interpolated. In most most cases, this is not an issue as emissivity parameterizations provide a sufficiently high accuracy \citep[see][]{perez-montero_ionized_2017,peimbert_nebular_2017}. However, in the case of the helium transitions, up to $5\%$ discrepancy was observed for some temperature-density regions. In this study, this handicap was removed by including the \textit{RegularGridInterpolator} function from the \textsc{exoplanet} library by \cite{dan_foreman-mackey_dfm/exoplanet:_2019}. This function enables the interpolation on a regular tensor with an arbitrary number of dimensions. The inclusion of this function increases simulation time up to three times in our synthetic test cases. However, it makes possible a more faithful characterisation of the emissivities computation and the inclusion of more emission lines. The atomic data considered in the chemical analysis can be found in Table \ref{tab:atomic-data} within the appendix

\end{itemize}

\subsection{Characterization of the ionizing radiation}

\begin{table*}
\caption{\label{tab:atomic-data}Atomic data references for the emission lines considered in the chemical analysis.}
\centering{\begin{tabular}{ccc}
\hline 
Ion & \multicolumn{2}{c}{Atomic data}\tabularnewline
\hline 
\hline 
$H$ & \multicolumn{2}{c}{\citet{storey_recombination_1995}}\tabularnewline
\hline 
$He$ & \multicolumn{2}{c}{\citet{porter_improved_2012}}\tabularnewline
\hline 
$He^{+}$ & \multicolumn{2}{c}{\citet{storey_recombination_1995}}\tabularnewline
\hline 
\hline 
Ion & Collision Strengths & Transition probabilities\tabularnewline
\hline 
\hline 
$O^{+}$ & \citet{pradhan_[o_2006,tayal_oscillator_2007} & \citet{zeippen_transition_1982,wiese_atomic_1996}\tabularnewline
\cline{1-3} 
$S^{+}$ & \citet{tayal_breitpauli_2010} & \citet{podobedova_critically_2009} \tabularnewline
\hline 
$O^{+2}$ & \citet{aggarwal_vizier_2000} & \citet{storey_theoretical_2000,wiese_atomic_1996}\tabularnewline
\cline{1-3} 
$N^{+}$ & \citet{tayal_electron_2011} & \citet{wiese_atomic_1996,galavis_atomic_1997}\tabularnewline
\cline{1-3} 
$S^{+2}$ & \citet{hudson_collision_2012} & \citet{podobedova_critically_2009}\tabularnewline
\cline{1-3} 
$S^{+3}$ & \citet{tayal_electron_2000} & \citet{dufton_s_1982,johnson_atomic_1986}\tabularnewline
\cline{1-3} 
$Ar^{+2}$ & \citet{galavis_atomic_1995} & \citet{kaufman_forbidden_1986,galavis_atomic_1995}\tabularnewline
\cline{1-3} 
$Ar^{+3}$ & \citet{ramsbottom_effective_1997} & \citet{mendoza_transition_1982}\tabularnewline
\hline 
\end{tabular}
}
\end{table*}

Recently, \citet{perez-montero_photon_2020} delved into the physics of HeII-emitters. They proposed a Bayesian-like algorithm to sample photoionization grids, in which the $[OII]3727,3729\AA$, $[OIII]5007\AA$, $[SII]6717,6731\AA$, $[SIII]9069\AA$, $HeI4471\AA$, $HeI5876\AA$ and $HeII4686\AA$ line fluxes are a function of the oxygen abundance, the effective cluster temperature $(T_{eff})$ and the ionization parameter $(log(U))$. The former term quantifies the temperature of a single source responsible for the ionizing radiation. The latter term represents a dimensionless ratio of the ionization photon density to the electron density. Different photoionization grids were generated
using \textsc{Cloudy v17} \citep[see][]{ferland_2017_2017} to map $T_{eff}$ according to different SEDs.
For this work, we consider a blackbody with $T_{eff}$ ranging from $30kK$ to $90kK$. According to \citet{perez-montero_photon_2020}, these grids are the only ones with high enough $T_{eff}$ values, to generate the $HeII4686\AA$ fluxes in the control sample. The $log(U)$ value goes from -4.0 to -1.5. Finally, the ionized gas metallicity is parameterized according to the oxygen content, where $12+log(\nicefrac{O}{H})$ increases from 7.1 to 8.9.  Additionally, the algorithm photoionization grids include both a spherical and plane-parallel geometries as defined by the one-dimensional Cloudy model. This algorithm is named \textsc{HII-CHI-mistry-Teff} and it shares a similar sampling procedure to the \textsc{HII-CHI-mistry} algorithm presented in \cite{perez-montero_deriving_2014} and \citet{perez-montero_using_2017}: For the selected model grid, the algorithm starts by slicing the grid given the input $\nicefrac{O}{H}$ abundance. Afterwards, the algorithm loops through the theoretical $HeII4686\AA$ fluxes. At non-zero values, the code compares the observed $\nicefrac{[OII]}{[OIII]}$ or $\nicefrac{[SII]}{[SIII]}$ ratios (the one provided by the user) and helium fluxes against the theoretical values. This results in a set of $\chi^{2}$ weights. This process is repeated for an input number of iterations where the line fluxes are randomly generated in a standard Monte Carlo process: A normal distribution centred at the observed flux and a standard deviation given by the line flux uncertainty. The final $T_{eff}$ and $log(U)$ values are computed from $\chi^{2}$-weighted traces.

As a novelty, this work introduces a \textsc{HII-CHI-mistry-Teff} algorithm written with neural networks. This model adheres to a Bayesian paradigm following the same structure described in section \ref{chemical_analysis}: The algorithm starts defining a prior for $T_{eff}$ and $log(U)$ (see Table \ref{tab:Priors-and-likelihood}). From the input oxygen abundance, an additional prior is defined with a normal distribution with $\mu=\nicefrac{O}{H}$ and $\sigma=\sigma_{\nicefrac{O}{H}}$. This informative prior is introduced to account for the uncertainty in the elemental abundance measurement. Afterwards, the $T_{eff}-log(U)-\nicefrac{O}{H}$ grid is interpolated using the algorithm from \citet{dan_foreman-mackey_dfm/exoplanet:_2019}. Finally, the line intensities are extinguished according to the input $c(H\beta)$ coefficient prior their evaluation against the observed values. This likelihood is defined using the same schemed as in the chemical model (see Table \ref{tab:Priors-and-likelihood}). 

This new design can be integrated with in the chemical model described in section \ref{chemical_analysis}. In this case, rather than using a constant oxygen abundance to interpolate the photoionization grids, this is calculated from the sampled $O^{+}$ and $O^{2+}$ values. Similarly, the emission relative line intensities are extinguished at each iteration form the currently proposed $c(H\beta)$. This brings the number of dimensions for the GPs sample spectra fitting to 16. In the Bayesian \textsc{HII-CHI-mistry-Teff} and the Bayesian Direct method + \textsc{HII-CHI-mistry-Teff} only the plane parallel geometries are considered. This selection is partially imposed by the fact that these are the only regular grids included in \textsc{HII-CHI-mistry-Teff}. Learning from this work conclusions, future research will be focused on generating tailored photoionization grids for a Bayesian sampler. This is, though, beyond the scope of the current work.

\section{Discussion} \label{discussion}

A careful inspection of the sample continua (see Fig. \ref{fig:sample_spectra} lower panels in logarithmic scale) confirms no evidence of a blue $(\lambda\sim4650\AA)$ or red bump $(\lambda\sim5808\AA)$. Unlike in the \citetalias{amorin_star_2012} sample, the current spectra do not show WR features. Moreover, none of the objects display noticeable absorption features, such as the $MgI\lambda\lambda5167,5173\AA$ lines. All three objects, however, exhibit nebular $HeII4686\AA$ emission, which demonstrates the existence of a very hard ionization field \citep[see][]{kehrig_extended_2015,kehrig_extended_2018}. Due to the sample redshift, it can be noticed that several telluric features are contaminating the optical range. In GP030321, a sky band has completely swallowed the $[NII]6548\AA$ emission. Fortunately, it does not affect the $H\alpha$ flux. Moreover, in GP101157 a telluric absorption has a detrimental impact on the $HeI6678\AA$ line. Consequently, it is excluded from the chemical analysis. Finally, from the lower panels in Fig. \ref{fig:sample_spectra} the Balmer jump $(\lambda\sim3646\AA)$ can be easily appreciated on GP101157 and GP121903. The smaller jump in GP030321 could be explained by an older star forming region.

\subsection{Analysis of the synthesis spectral continuum fitting} \label{stellar-discussion}

\begin{figure}
\includegraphics[width=1.0\columnwidth]{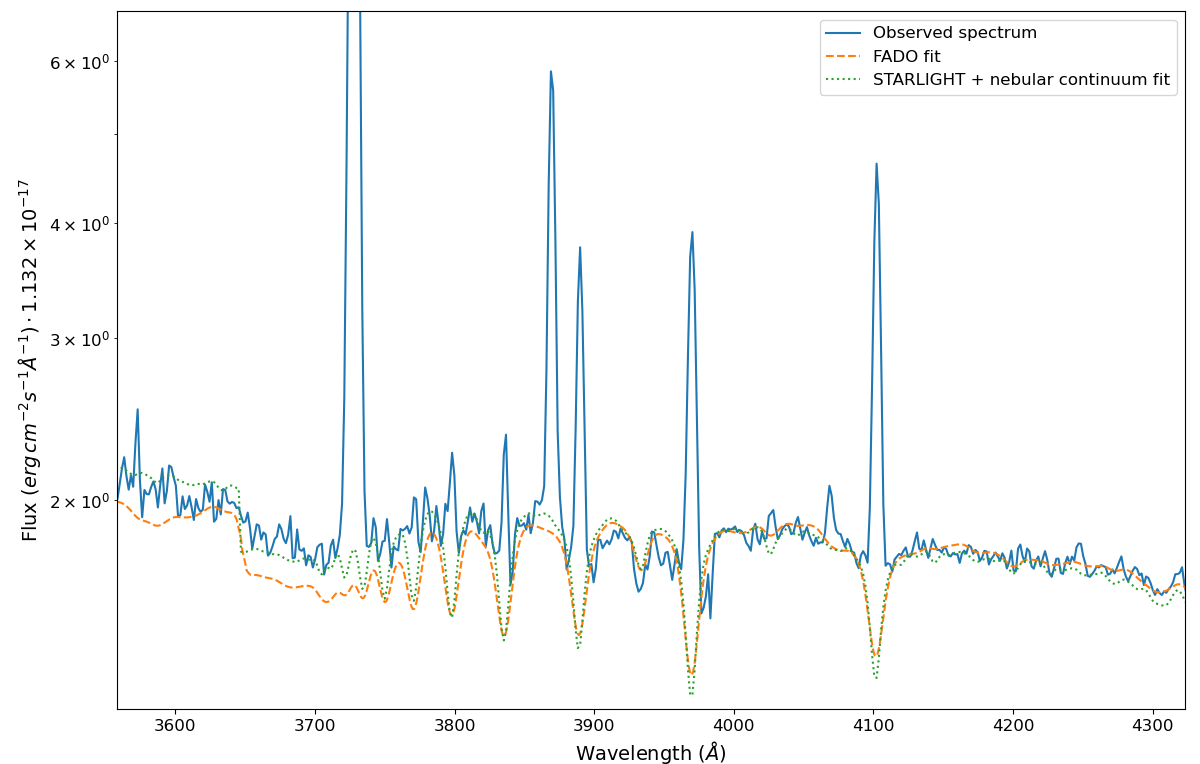}
\caption{\label{fig:SSP-comparison}SSP synthesis output continua comparison for galaxy GP232539}
\end{figure}

\begin{table*}
\caption{\label{tab:FADO-sample-output} FADO results for the SSP synthesis for the GP sample. Column (1) displays the object name. Column (2) shows fit $\chi^{2}$ coefficient (top) and the nebular component luminosity percentage (bottom). Column (3) shows the SSPs mean metallicity weighted by light (top) and mass (bottom). Column (4) displays the average age, in logarithmic scale, weighted by light (top) and mass (bottom). Column (5) displays the logarithmic mass ever formed: All populations total (bottom) and the fraction above 1Myr. Column (6) displays the logarithmic mass currently available: All populations total (bottom) and the fraction above 1Myr. Finally, column (7) displays the stellar (top) and nebular (bottom) extinction. The uncertainties predicted by FADO are $\Delta z\approx0.005$, $\Delta log(t)\approx0.005$, $\Delta log(M)\approx0.001$ and $\Delta c(H\beta)\approx0.01$ }
\centering{{\begin{tabu}{lcccccc}%
\hline%
Galaxy&$\chi^2$&$\left\langle z_{M}\right\rangle$&$\left\langle log(t_{L}) \right\rangle$&$log(M_{e})$&$log(M_{c})$&$A_{V,\,stellar}$\\%
\hline%
GP030321&$17.79$&$0.029$&$7.301$&$9.094$&$8.864$&$0.769$\\%
GP101157&$20.76$&$0.012$&$7.832$&$8.752$&$8.591$&$0.971$\\%
GP121903&$5.072$&$0.002$&$8.103$&$9.497$&$9.204$&$-0.287$\\%
GP004054&$2.806$&$0.001$&$6.880$&$8.113$&$8.039$&$0.954$\\%
GP113303&$3.689$&$0.049$&$7.545$&$9.458$&$9.167$&$-0.419$\\%
GP232539&$16.79$&$0.003$&$8.039$&$9.522$&$9.250$&$0.382$\\%
\hline%
Galaxy&$L_{nebular}\,(\%)$&$\left\langle z_{L}\right\rangle $&$\left\langle log(t_{M}) \right\rangle $&$log(M^{> 1Myr}_{e})$&$log(M^{> 1Myr}_{c})$&$A_{V,\,nebular}$\\%
\hline%
GP030321&$10.08\%$&$0.021$&$9.038$&$8.957$&$8.664$&$-0.050$\\%
GP101157&$27.57\%$&$0.017$&$8.208$&$7.276$&$7.020$&$0.142$\\%
GP121903&$23.28\%$&$0.003$&$9.802$&$9.483$&$9.179$&$0.290$\\%
GP004054&$17.19\%$&$0.002$&$7.227$&$0.000$&$0.000$&$0.436$\\%
GP113303&$10.92\%$&$0.021$&$9.967$&$9.454$&$9.160$&$0.100$\\%
GP232539&$7.40\%$&$0.011$&$9.481$&$9.473$&$9.181$&$0.720$\\%
\hline%
\end{tabu}}} 
\end{table*}

\begin{figure*}
\includegraphics[width=1.0\textwidth]{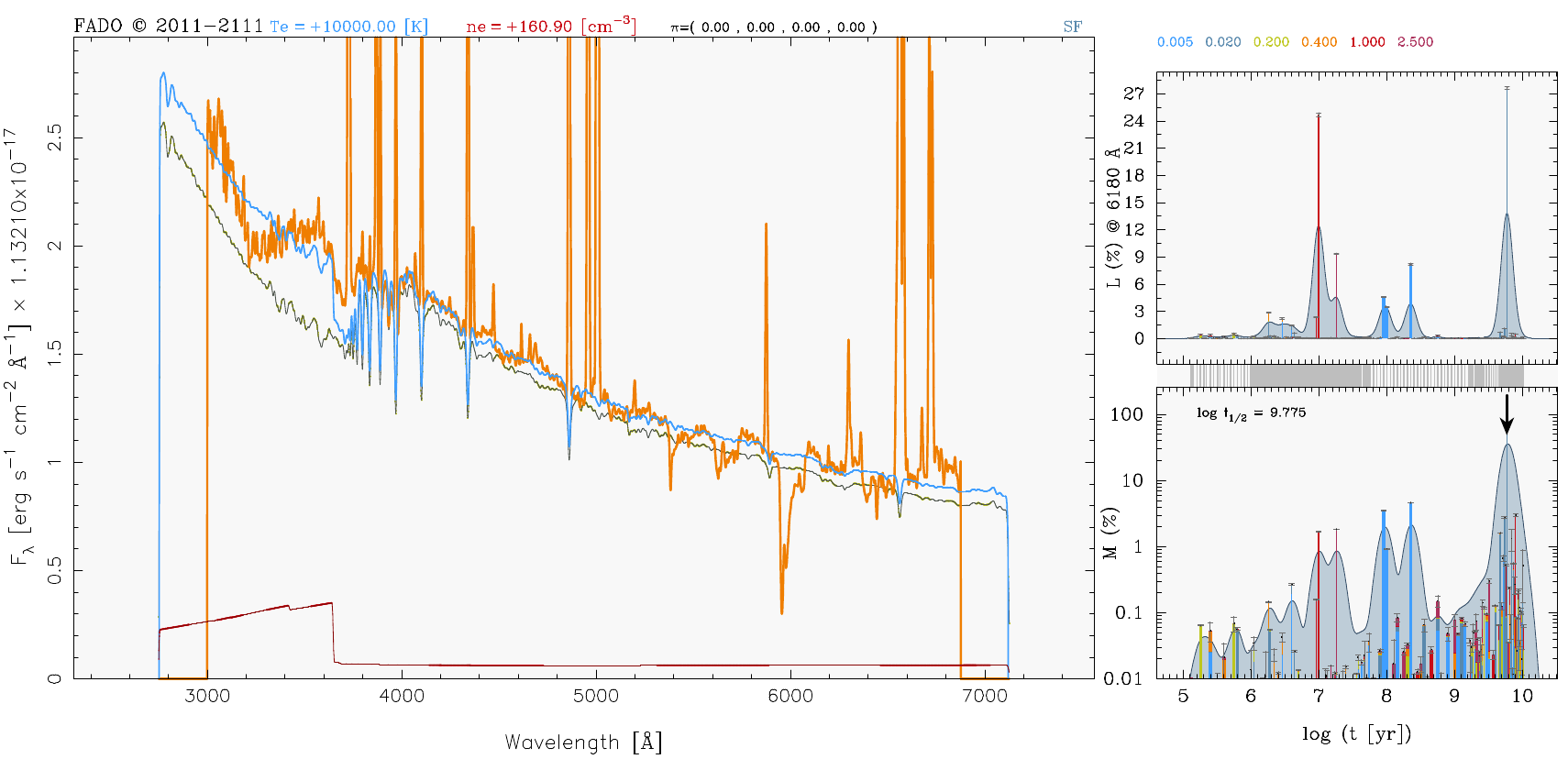}
\caption{\label{fig:fado-infograph}\textsc{FADO} continua fit for the galaxy GP232539 (orange). The best fitting SED (blue) includes the stellar (black) and nebular (red) components. The histograms on the upper and lower right corners display the luminosity and mass fraction at the normalization wavelength $(6180\AA)$. The color-coding depicts the metallicity and the vertical bars the $\pm1\sigma$ uncertainties (the values can be found at the top right  top right corner). The grey bars in between both plots correspond to the ages for the stellar grid. The vertical arrow in the mass fraction diagram marks the age when 50\% of the present-day stellar mass has been in place. Same plots for the remaining galaxies can be found in the appendix}
\end{figure*}

Due to the intense emission line features, it is easy to ignore the gas contribution to the continuum. For example,  \citet{reines_emerging_2008} noticed an excess on the Hubble U and I bands observation of massive stellar clusters, which could not be explained by the stellar component. In posterior work, \citet{reines_importance_2009} succeeded to explain this excess from the nebular continuum and the sulfur emission at $[SII]9069,9531\AA$. As we delve into higher redshifts and more intense star-forming systems, it is essential to take into account this gas contribution \citep[see][]{schaerer_impact_2009,amorin_extreme_2015}. The two SSPs synthesis techniques we apply in this analysis include the nebular continuum computation. The first one follows the iterative analysis presented in \citet{fernandez_determination_2018} for the high accuracy determination of the primordial helium abundance from local star forming galaxies. The second methodology involves the application of \textsc{FADO} by \citet{gomes_fitting_2017}.

Fig. \ref{fig:SSP-comparison} shows both techniques (see sec. \ref{sec:nebular_calc} and sec. \ref{sec:stellar-calc}) output against the GP spectrum. This fitting corresponds to the galaxy GP232559 from the \citetalias{amorin_star_2012} sample, however, similar conclusions are drawn from the six spectra fits. Both techniques show a good visual match with the observations, even though the output $\chi^{2}$ coefficients can reach double digits (see Table \ref{tab:FADO-sample-output}). The largest discrepancies are found at the blue end, in the Balmer jump proximity. The iterative \textsc{Starlight} + nebular computation analysis displays a better agreement with the observed continuum for $\lambda>3646\AA$. In contrast, this technique overestimates the continuum flux for $\lambda<3646\AA$ region. The opposite pattern is observed on the \textsc{FADO} fitting. A couple of causes can contribute to this effect: In the \textsc{FADO} fits, the region $3600\AA<\lambda<3900\AA$ is completely masked, while in the \textsc{Starlight} only the emission line regions are masked. A large mask can make it hard to for the algorithm to fit the continuum, due to the limited number of data points. Nonetheless, it may be argued that even if we cannot see the emission lines from Balmer transitions above $HI_{Balmer}>16$ this emission rich region should still be masked. Another cause behind this discrepancy might be on the Balmer jump height calculation. The free-bound continuum features have a sensitive, inversely proportional relation to the hydrogen temperature \citep[see][]{zhang_electron_2004}. In the case of \textsc{FADO}, the $T_{[OIII]}$ temperature is used as a proxy to compute the nebular continua, while in the other approach, we use $T_{[OII]}$, leading to a higher jump height. According to the relation by \citet{liu_ngc_2000} a 10\% variation on the Balmer jump height measurement leads to almost a $1,000K$ discrepancy on the hydrogen temperature. The work by \citet{guseva_balmer_2006,guseva_balmer_2007} on low metallicity HII regions concluded that there was a good agreement on $T_{[OIII]}$ and $T_{HI}$ within the $O^{2+}$ zone. However, the authors warned that the $T_{HI}$ temperature determination was more reliable in regions with larger contribution from the gas. This means a higher $EW(H\beta)$, and therefore, younger objects. In the work by \citet{garcia-rojas_analysis_2012}, despite the high-quality spectra from planetary nebulae, the author refrained from comparing the Paschen discontinuity temperature with emission line diagnostics due to the high uncertainty involved. From this comparison, the choice between low or high ionization temperature has a small impact on the resulting stellar continuum fit (against the option of neglecting the nebular component all together). However, due to these GPs rich emission features, there is a shortage of "blue" continuum. This means that the fixed Balmer jump height has a large weight, in a region where the input spectrum is mostly masked. Consequently, we may be properly detecting the youngest stars, but their mass/luminosity contribution is underestimated.

Regarding the stellar kinematics, in general, the \textsc{FADO} output predicts slightly weaker absorption features with larger stellar dispersion velocities. This discrepancy can be explained by the boundary condition in the \textsc{Starlight} fitting, where the kinematics must remain below the $\sigma_{[OIII]}$ value. It can be appreciated in Fig. \ref{fig:SSP-comparison} that GP spectra have almost imperceptible hydrogen absorptions. Due to this, most algorithms attempt to overestimate stellar dispersion velocity to better simulate the input "flat" spectrum. Consequently, the stellar dispersion velocity measured is actually the upper boundary condition for this parameter. In the literature, the uncertainty on the absorption features strength is a concern for high precision measurements of the helium abundance \citep[see][]{izotov_new_2014, aver_effects_2015, peimbert_primordial_2016}. In these works, it is commonly accepted that the impact on the hydrogen and helium is of the order of $Eqw \approx 2-0.5\AA $ respectively. These values were taken from synthetic stellar libraries such as those from \cite{delgado_synthetic_1999,delgado_synthetic_1999-1}. In the updated analysis by \citet{aver_improving_2020}, the authors quantified the hydrogen and helium line absorptions from the BPASS library \citep[see][and references therein]{stanway_re-evaluating_2018}. They proposed that this uniform ion absorption should be corrected by a relative scaling factor for each transition. In the case of the $H\alpha$, $H\gamma$ and $H\delta$ the scaling factor (relative to the $H\beta$ absorption) stands in the 0.964-0.930 range. In the case of the helium lines, the scaling factor (relative to the $HeI4471\AA$ absorption) varies within 0.346-1.4 range. Unlike the hydrogen lines, the helium absorptions strength shows no constant behaviour with the transition wavelength. 
The small discrepancies between both techniques can be explained by the previous arguments. At this point, however, we take the results from \textsc{FADO} as the most accurate. Moreover, given the larger SSPs library provides, it provides a more detailed characterization on SFH. In contrast, the output stellar continuum from the iterative approach is the one used for the chemical analysis. Since this methodology applies the $c(H\beta)$,  $\sigma_{[OIII]}$, $Te$, $y^+$ and $y^{2+}$ measured from the emission spectrum, it is deemed to provide a more tailored correction for the recombination lines.

Fig. \ref{fig:fado-infograph} displays the output from the SSPs synthesis for GP232539. The complete infographics for the current and  \citetalias{amorin_star_2012} sample can be found on the appendix. However, Table \ref{tab:FADO-sample-output} provides a summary with the main results. The sample rundown leads to the following conclusions: Both the current and \citetalias{amorin_star_2012} are confirmed to be dwarfish systems with $\overline{\left\langle log(M_e) \right\rangle}\approx9\,M_{\odot}$. As expected, the younger populations dominate the galaxy luminosity, $\overline{\left\langle log(t_L) \right\rangle}\approx7.6\,yr$, while the older populations dominate the mass, $\overline{\left\langle log(t_M) \right\rangle}\approx8.9\,yr$. As discussed in \citetalias{amorin_star_2012} the younger population peaks around $log(t) \approx 7\, yr$ while the older population remains at $log(t) > 9\, yr$. The new methodology confirms the presence of very young stars with $log(t) < 6\, yr$. These stars could not be properly identified in \citetalias{amorin_star_2012} due to the very  limited number of young stars in the SSP grid. Their light contribution, however, accounts only for a few percentage points and their combined mass one order below that. The output visual extinction coefficients have been converted to the relative logarithmic extinction coefficients for comparison's sake. In the case of GP101157 and GP121903 a good agreement is found with the nebular extinction from the results of the iterative analysis in Table \ref{tab:emission_fluxes_gp03032}. For GP030321, however, the extinction coefficient displays a negative value. Negative extinctions are also fitted for the stellar component of GP121903 and GP113303. This can be explained by the very low extinction encountered in these objects, making it difficult to anchor the parameter value. As in the case of \textsc{Starlight}, a non-negative extinction boundary must be imposed on these galaxies. Assuming the low uncertainty limits on the \textsc{FADO} measurements, the extinction between the stellar and nebular components are distinctively different. 

Focusing individually on the GPs stellar distribution, a few patterns emerge which must be highlighted:

\begin{itemize}
    \item The stellar masses measured for GP113303 and GP232539 display a very good agreement with the values presented in \citetalias{amorin_star_2012}. In the case of GP004054, however, the current value is a $13\%$ lower as the oldest stellar population ($log(t) > 8\, yr$)  is not detected in the SSP synthesis. In \citetalias{amorin_star_2012}, only 3 stellar populations (two young and one old) were detected in GP004054 fitting. In the present methodology, both GP004054 and GP113303 still display few SSPs (below the dozen), however the \textsc{FADO} fitting predicts a wider age range for the younger stellar populations. In  \citetalias{amorin_star_2012} both galaxies displayed a $\Delta log(t) \approx 6.8-7\, yr$ for both galaxies, while the current approach predicts a time interval $\Delta log(t) \approx 5-7.5\, yr$ for GP004054 and $\Delta log(t) \approx 6-7\, yr$ for GP113303. In general, few SSPs in the output continuum are due to an input spectra with small wavelength range. Due to the emission lines, the WR features, the sky bands and the noisy regions, GP004054 has irrefutably the smallest amount of data points in the SSP synthesis. Finally, some discrepancies between both results may be explained by the larger SSP library in the current work. 
    \item The application of the new methodology alongside the very fine stellar grid results in a more complex set of stellar distributions in the GP sample. In the case of the younger populations, the light and mass distributions show similar profiles. This is not the case, however, for the older stellar populations. On the one hand, the mass fraction histograms show a continuous distribution of SSPs, which peak around $log(t)\approx 8.5yr$ and even higher in some GPs. On the other hand, the light fraction in some GPs display a single SSP spiking above the rest. This is the case of GP232539 in Fig. \ref{fig:fado-infograph}. This is also the case of GP101157 and GP121903. The same behaviour could be attributed to GP004054 and GP113303, however, as discussed above, due to the limited number of SSPs fitted, few constrains can be derived from their SFH. In contrast, GP030321 despite having a light and mass fraction distributions very similar to GP232539 shows a light contribution from the older population below the $2\%$. The physical justification behind a single source dominating the luminosity, but not the mass, could be found on the nebular emission. However, since our methodology quantifies the contribution from the gas component, these results open two possible scenarios:
    
\end{itemize}

On the one hand, the calculated nebular continuum is underestimated. It is known that the gas contribution tends to make the stellar SED redder than expected \citep[see][]{krueger_optical_1995,papaderos_age_1998,izotov_green_2011}. In the current GP sample, the nebular continuum computed by FADO is 10.1\%, 27.6\% and 23.28\%. Similar values were found for the continuum level of $H\alpha$ and $H\beta$ for the GPs in \citetalias{amorin_star_2012}. This work included an analysis on the Hubble photometry for the GP sample. It confirmed that a LSB (Low Surface Brightness) periphery was appreciated for all GPs which could be associated to the underlying galaxy. The authors, however, included a warning that this component could be enhanced, or dominated by the nebular emission. This phenomenon would decrease the absolute magnitude of the host galaxy by $0.75 - 1\,mag$. Recently, \citet{clarke_old_2020} accomplished a similar analysis, carefully selecting a GP sample, whose Hubble photometry avoided the oxygen and hydrogen strongest lines. They also concluded that the LSB envelope is compatible with evolved hosts for BCDs and the nebular continuum contributes to the redder colours with $B-I\approx 0.76$. Nonetheless, given the multi-cluster nature of some GPs, it cannot be excluded that the young stars and ionized gas have a mismatching spatial distribution. This scenario was investigated by \citet{papaderos_i_2012} for IZw18, where the nebular emission envelope supports a non-cospatial scenario with the underlying ionizing and/or the non-ionizing stellar background. Even though this galaxy might be an extreme BCDs in the local universe, its properties might as well be ordinary at higher redshifts. Assuming a standard geometry could bring many biases to the measurements, such as the ones observed in this sample. Recently, the work by \citet{leja_older_2019} concluded that the 3D-HST, 0.5 < z < 2.5 catalogues have systematically older SFHs using the flexible SFH algorithm \textsc{Prospector-$\alpha$}. In these objects, the largest offsets are caused by the $t > 100 Myr$ stars. These results motivate more and deeper observations for a better characterization of extreme star forming systems.

On the other hand, the stellar population synthesis may be overestimating the luminosity from the older stellar populations. A self-consistent algorithm such as FADO, not only fits the best SFH but also the nebular features: the hydrogen continuum discontinuities, the hydrogen line luminosities and their equivalent widths. A large continuum photon escape fraction may have a detriment effect on the emission gas features. As fewer younger stars are necessary to explain the ionized gas, the algorithm balances the continuum fit with a larger contribution from the older stellar populations. These complex scenarios with have been explored by  are left to investigate for a forthcoming paper (Papaderos et al, in preparation). 

\subsection{Analysis of the resulting gas physical properties and chemical abundances} \label{emission-discussion}

\begin{table}
\caption{\label{tab:fit_results}Resulting gas physical parameters and chemical abundances from the    multidimensional emission-line fitting in the three analyzed GP galaxies.}
\centering{{\tiny\begin{tabu}{lccc}%
\hline%
Parameter&GP030321&GP101157&GP121903\\%
\hline%
$n_{e}(cm^{-3})$&$24\pm14$&$500\pm52$&$547\pm63$\\%
$T_{high}(K)$&$15700\pm200$&$14600\pm100$&$15700\pm200$\\%
$c(H\beta)$&$0.71\pm0.03$&$0.80\pm0.03$&$0.03\pm0.02$\\%
$\frac{Ar^{2+}}{H^{+}}$&$5.37\pm0.03$&$5.68\pm0.02$&$5.62\pm0.03$\\%
$\frac{Ar^{3+}}{H^{+}}$&$4.82\pm0.05$&$5.18\pm0.01$&$5.10\pm0.02$\\%
$\frac{Fe^{2+}}{H^{+}}$&$5.04\pm0.05$&$5.18\pm0.02$&$5.13\pm0.02$\\%
$\frac{N^{+}}{H^{+}}$&$6.03\pm0.06$&$6.19\pm0.11$&$6.48\pm0.05$\\%
$\frac{Ne^{2+}}{H^{+}}$&$7.13\pm0.02$&$7.23\pm0.01$&$7.04\pm0.02$\\%
$\frac{O^{+}}{H^{+}}$&$6.88\pm0.04$&$7.59\pm0.05$&$7.44\pm0.05$\\%
$\frac{O^{2+}}{H^{+}}$&$7.70\pm0.02$&$7.85\pm0.01$&$7.77\pm0.02$\\%
$\frac{S^{+}}{H^{+}}$&$5.36\pm0.03$&$5.56\pm0.03$&$5.49\pm0.03$\\%
$\frac{S^{2+}}{H^{+}}$&$6.02\pm0.06$&$6.61\pm0.04$&$6.43\pm0.05$\\%
$y^{+}$&$0.12\pm0.01$&$0.08\pm0.01$&$0.08\pm0.02$\\%
$y^{2+}$&$\num{1.9e-03}\pm\num{8e-05}$&$\num{1.3e-03}\pm\num{5e-05}$&$\num{2.2e-03}\pm\num{4e-05}$\\%
\hline%
\end{tabu}}}
\end{table}

\begin{figure}
\includegraphics[width=1.0\columnwidth]{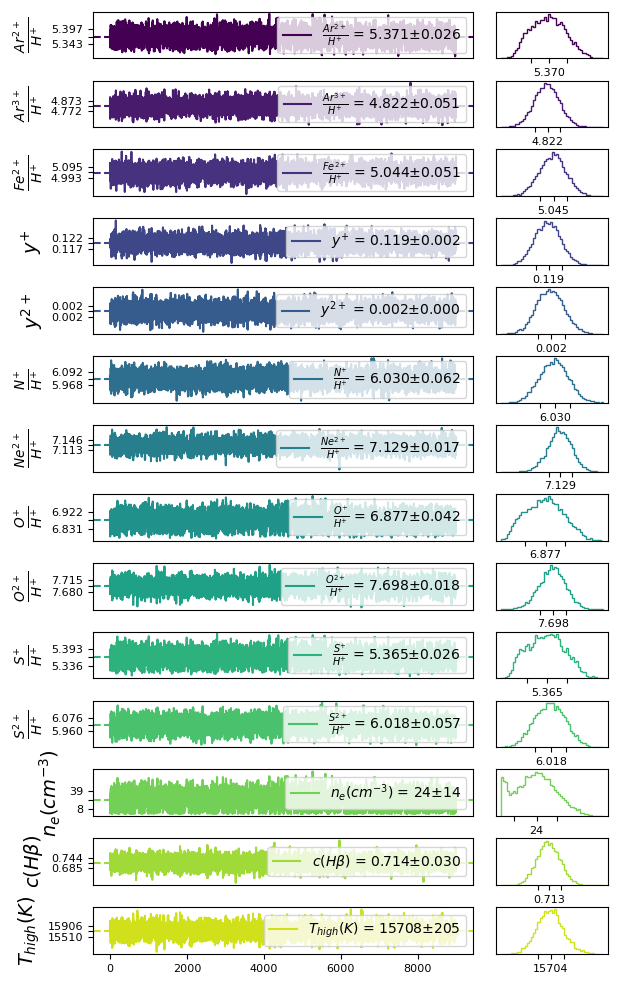}
\caption{\label{fig:GP030321_outputFitPlot}Multi-dimensional fit plot for the chemical model of GP030321}
\end{figure}

\begin{figure}
\includegraphics[width=1.0\columnwidth]{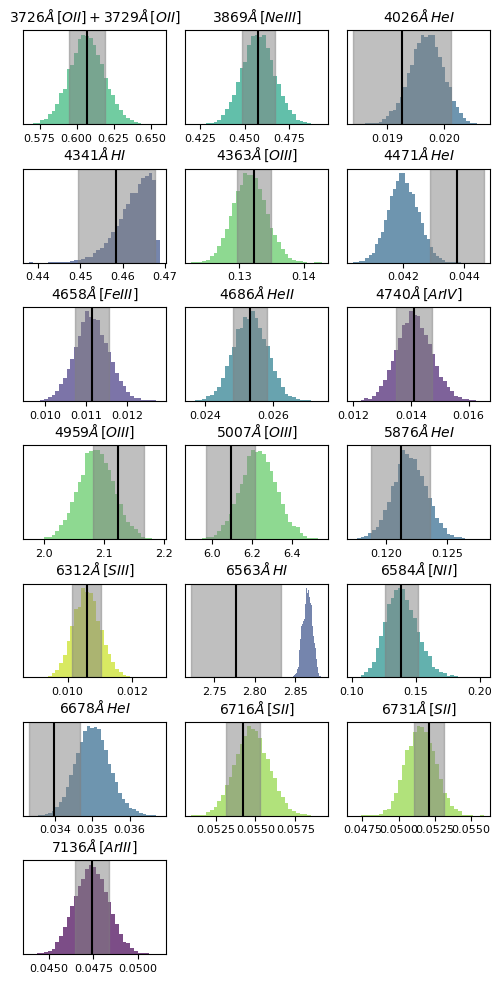}
\caption{\label{fig:GP121903_flux_FitPlot} Histogram grid with the synthetic fluxes for the emission lines fitted for GP121903. The distributions are colour coded (in the electronic version) according to the ion generating the photons. The vertical black line and grey shaded area correspond to the observed line fluxes with their uncertainty}
\end{figure}

The data in Table \ref{tab:fit_results} shows the results from the multidimensional analysis. For GP030321, the same data is displayed graphically in Fig. \ref{fig:GP030321_outputFitPlot}. As it was explained in \citet{fernandez_bayesian_2019}, these diagrams provide a quick diagnosis on the Monte Carlo algorithm fitting and convergence quality: On the left-hand side, the white noise-like plot represent the parameter trace at each of the 9000 simulation steps. As long as no divergences are encountered, on any parameter, it can be stated that the sampling has reached convergence. Moreover, these traces consist in three distinctive simulations (running three cores in parallel). A seamless juncture between all runs, as it is the case in the traces in Fig. \ref{fig:GP030321_outputFitPlot}, means that the initial conditions are not affecting the measurements. On the right-hand side, the same traces are plotted as a histogram. This is the Bayesian posterior, and consequently, our measurement.

\begin{table}
\caption{\label{tab:elemental-abundances} Elemental abundances and chemical parameters computed from the results of the Bayesian direct method model.}
\centering{\begin{tabu}{cccc}%
\hline%
Parameter&GP030321&GP101157&GP121903\\%
\hline%
$\nicefrac{O}{H}$&$7.76\pm0.02$&$8.04\pm0.02$&$7.94\pm0.02$\\%
$\nicefrac{N}{H}$&$6.91\pm0.06$&$6.64\pm0.11$&$6.98\pm0.04$\\%
$\nicefrac{S}{H}$&$6.22\pm0.05$&$6.79\pm0.04$&$6.60\pm0.04$\\%
$\nicefrac{He}{H}$&$0.12\pm0.00$&$0.08\pm0.00$&$0.09\pm0.00$\\%
\hline%
$log(\nicefrac{N}{O})$&$-0.85\pm0.06$&$-1.40\pm0.11$&$-0.96\pm0.04$\\%
$ICF\left(S^{3+}\right)$&$1.30\pm0.03$&$1.38\pm0.02$&$1.36\pm0.03$\\%
\hline%
\end{tabu}

}
\end{table}

An additional quality test is available in Fig. \ref{fig:GP121903_flux_FitPlot}. This grid shows the fitting theoretical fluxes calculated from Eq. \ref{eq:fluxFormula} against the observational ones. At this point, the reader is encouraged to check  the online support material, where Fig. \ref{fig:GP030321_outputFitPlot} and Fig. \ref{fig:GP121903_flux_FitPlot} can be found for the remaining two objects. It can be stated that in general, the fits agree very well with the observational data. Nonetheless, the few discrepancies should be explained:

\begin{itemize}
    \item $c(H\beta)$ parameter: In the simulation of GP121903, the logarithmic extinction posterior displays a bounded distribution. This is explained by the parameter prior, a half-Cauchy, distribution which forbids negative parameter values. In this galaxy with exceptionally low extinction, as the simulation tries to sample negative coefficients the measurement ends-up being constrained. This pattern is also observed in the $H\gamma$ flux posterior plot for GP121903. Consequently, this line is the one pushing towards negative extinction coefficients. The likely cause is the correction for the underlying stellar absorption which not being sufficiently accurate. A solution for this issue could be a prior which simulates an asymptotic behaviour as $c(H\beta)$ approaches zero. Nonetheless, since the dust extinction is very small the impact from the hydrogen lines miss-match has a very small impact on the rest of the parameter space.
   
    \item $n_e$ parameter: In the simulation of GP030321 the density posterior appears to be double peaked. One value around $n_e = 25cm^{-3}$ and the second peak at zero. As in the previous case, this is caused by the simulation priors, which have been designed to avoid non-physical values. Moreover, the emissivity grids are constrained to a density range of $n_e=1-800cm^{-3}$. A traditional density calculation using the $ne[SII]$ parameterization by \citep[see][]{perez-montero_ionized_2017} results in a negative density value. Trying to explain these results, a careful inspection of the line profiles may lead to the conclusion of a wider or secondary profile in the $[SII]6716\AA$ line. This behaviour has been recently analysed by \citet{hogarth_chemodynamics_2020} in GPs with clear signatures of outflows and turbulence. This could explain the non-physical emissivities derived for this object. However, a secondary component would also be present on the recombination and collisionally excited auroral lines. This is not the case for any galaxy of this sample. Therefore, we can dismiss an outflow. A review on the Ultraviolet and Visual Echelle Spectrograph (UVES) sky map by \citet{hanuschik_flux-calibrated_2003} confirms intense sky lines at 7821.5190$\AA$ and 7841.2847$\AA$. In GP030321 the $[SII]6717,6731\AA$ doublet is observed blended in the 7824-7841$\AA$ range. Consequently, the second component is heavily contaminated by the sky. This can explain the doublet profile difference after the background subtraction task. The direct impact of this bimodal distribution on the emissivity computation is negligible for the observed transitions at $n_{e} < 100cm^{-3}$ regime. However, there is an indirect impact due to the temperature value of $T[OII]$ which from eq. \ref{eq:TOIII-TSIII-relation} depends on $n_{e}[SII]$. Indeed, wider priors (i.e. larger uncertainty) is observed on the low ionization abundances. This can be explained by the density distribution. Nonetheless, $T[OII]$ also depends on $T[OIII] (T_{high})$, which as we may see in Fig.\ref{fig:GP030321_outputFitPlot} has a normal posterior. This means that the temperature fit is not affected by the density uncertainty. Therefore, the fitting is behaving as expected: larger uncertainty only on the parameters which depend on observables with higher error.
    
    \item $HeI$ fluxes: In the case of GP030321 and GP101157 large discrepancies ($\sim20-40\%$) between the theoretical and observed fluxes are found for the helium lines. As it was discussed in \citet{fernandez_determination_2018,fernandez_bayesian_2019} the observed HeI photons are affected by several phenomena whose characterization is challenging. The main one is the absorption for the underlying stellar population whose accuracy on the absorption features strength cannot be easily quantified. Moreover, no error propagation takes place between the stellar population synthesis and its output removal from the observed spectrum. This means that the uncertainty is underestimated in these fits, independently of the line wavelength. This phenomenon can also explain the poor fitting observed in some cases for the smaller wavelength HI transitions. 
\end{itemize}

\begin{table*}
\caption{\label{tab:photoIonization-param-comparison} Comparison between the photoionisation grid fitting described in the text for each galaxy.}
\centering{\begin{tabu}{lccccc}%
\hline%
Galaxy&Parameter&\makecell{\textsc{HII-CHI-mistry: } \\ Spherical}&\makecell{\textsc{HII-CHI-mistry: } \\ Plane-Parallel}&\makecell{Bayesian \textsc{HII-CHI-mistry: } \\ Plane-Parallel}&\makecell{Bayesian \\ Direct Method + \textsc{HII-CHI-mistry} \\ Plane-Parallel}\\%
\hline%
\multirow{2}{*}{GP030321}&$T_{eff}$&$57570\pm2982$&$62083\pm4362$&$75141\pm527$&$74054\pm563$\\%
&$log(U)$&$-2.09\pm0.27$&$-2.14\pm0.27$&$-2.59\pm0.01$&$-2.56\pm0.01$\\%
\multirow{2}{*}{GP101157}&$T_{eff}$&$56629\pm3151$&$60630\pm2821$&$64688\pm333$&$63997\pm391$\\%
&$log(U)$&$-2.00\pm0.28$&$-2.01\pm0.24$&$-2.13\pm0.01$&$-1.97\pm0.02$\\%
\multirow{2}{*}{GP121903}&$T_{eff}$&$56640\pm2077$&$61445\pm1784$&$68423\pm215$&$70021\pm125$\\%
&$log(U)$&$-1.76\pm0.19$&$-1.91\pm0.17$&$-1.85\pm0.01$&$-2.05\pm0.01$\\%
\hline%
\end{tabu}}
\end{table*}

\begin{table*}
\caption{\label{tab:photoIonization-flux-comparison} Observational fluxes against the synthetic fluxes fitted via the techniques described in the text for GP101157. The percentages correspond to the difference with the observed values in column}
\centering{{\tiny\begin{tabu}{lcccccc}%
\hline%
Line ID&Observed flux&\makecell{Bayesian: \\ Direct method}&\makecell{\textsc{HII-CHI-mistry: } \\ Spherical}&\makecell{\textsc{HII-CHI-mistry: } \\ Plane-Parallel}&\makecell{Bayesian \textsc{HII-CHI-mistry: } \\ Plane-Parallel}&\makecell{Bayesian \\ Direct Method + \textsc{HII-CHI-mistry} \\ Plane-Parallel}\\%
\hline%
$3726\AA\,[OII]+3729\AA\,[OII]$&$0.487\pm0.001$&$0.487\pm0.010$ $(0.0\%)$&$0.492\pm0.210$ $(1.0\%)$&$0.456\pm0.169$ $(-6.8\%)$&$0.480\pm0.006$ $(-1.5\%)$&$0.485\pm0.010$ $(-0.3\%)$\\%
$3869\AA\,[NeIII]$&$0.348\pm0.001$&$0.348\pm0.007$ $(0.0\%)$&-&-&-&$0.348\pm0.007$ $(0.0\%)$\\%
$4026\AA\,HeI$&$0.012\pm0.000$&$0.012\pm0.000$ $(2.2\%)$&-&-&-&$0.014\pm0.000$ $(13.1\%)$\\%
$4102\AA\,HI$&$0.179\pm0.001$&$0.170\pm0.003$ $(-5.2\%)$&-&-&-&$0.215\pm0.002$ $(16.8\%)$\\%
$4341\AA\,HI$&$0.320\pm0.001$&$0.351\pm0.004$ $(8.8\%)$&-&-&-&$0.412\pm0.003$ $(22.3\%)$\\%
$4363\AA\,[OIII]$&$0.086\pm0.001$&$0.087\pm0.002$ $(1.7\%)$&-&-&-&$0.089\pm0.002$ $(3.3\%)$\\%
$4471\AA\,HeI$&$0.024\pm0.000$&$0.033\pm0.000$ $(27.0\%)$&$0.017\pm0.000$ $(-36.9\%)$&$0.017\pm0.000$ $(-38.3\%)$&$0.017\pm0.000$ $(-37.7\%)$&$0.032\pm0.000$ $(25.3\%)$\\%
$4658\AA\,[FeIII]$&$0.010\pm0.000$&$0.010\pm0.000$ $(0.1\%)$&-&-&-&$0.010\pm0.000$ $(0.0\%)$\\%
$4686\AA\,HeII$&$0.014\pm0.000$&$0.014\pm0.000$ $(0.1\%)$&$0.006\pm0.003$ $(-147.2\%)$&$0.009\pm0.003$ $(-61.1\%)$&$0.012\pm0.000$ $(-22.2\%)$&$0.014\pm0.000$ $(0.1\%)$\\%
$4740\AA\,[ArIV]$&$0.013\pm0.000$&$0.013\pm0.000$ $(0.0\%)$&-&-&-&$0.013\pm0.000$ $(0.0\%)$\\%
$4959\AA\,[OIII]$&$2.264\pm0.002$&$2.187\pm0.028$ $(-3.5\%)$&-&-&-&$2.171\pm0.027$ $(-4.3\%)$\\%
$5007\AA\,[OIII]$&$6.553\pm0.004$&$6.672\pm0.088$ $(1.8\%)$&$6.848\pm1.518$ $(4.3\%)$&$7.283\pm1.489$ $(10.0\%)$&$7.113\pm0.038$ $(7.9\%)$&$6.541\pm0.082$ $(-0.2\%)$\\%
$5876\AA\,HeI$&$0.124\pm0.001$&$0.090\pm0.001$ $(-37.9\%)$&$0.043\pm0.001$ $(-189.9\%)$&$0.042\pm0.001$ $(-193.2\%)$&$0.043\pm0.000$ $(-191.3\%)$&$0.089\pm0.001$ $(-40.0\%)$\\%
$6312\AA\,[SIII]$&$0.015\pm0.000$&$0.015\pm0.000$ $(0.1\%)$&-&-&-&$0.015\pm0.000$ $(0.1\%)$\\%
$6563\AA\,HI$&$2.795\pm0.027$&$2.865\pm0.005$ $(2.5\%)$&-&-&-&$2.869\pm0.004$ $(2.6\%)$\\%
$6584\AA\,[NII]$&$0.072\pm0.019$&$0.074\pm0.018$ $(2.7\%)$&-&-&-&$0.074\pm0.018$ $(2.8\%)$\\%
$6716\AA\,[SII]$&$0.073\pm0.001$&$0.069\pm0.001$ $(-5.0\%)$&-&-&-&$0.070\pm0.001$ $(-4.6\%)$\\%
$6731\AA\,[SII]$&$0.061\pm0.001$&$0.064\pm0.001$ $(4.8\%)$&-&-&-&$0.064\pm0.001$ $(4.4\%)$\\%
$6716\AA\,[SII]+6731\AA\,[SII]$&$0.134\pm0.001$&-&$0.112\pm0.053$ $(-19.8\%)$&$0.113\pm0.046$ $(-18.8\%)$&$0.138\pm0.002$ $(2.6\%)$&-\\%
$7136\AA\,[ArIII]$&$0.066\pm0.001$&$0.066\pm0.001$ $(0.0\%)$&-&-&-&$0.066\pm0.001$ $(0.0\%)$\\%
\hline%
\end{tabu}}}
\end{table*}

\begin{table}
\caption{\label{tab:fit_results_large}Chemical analysis results from the Direct method + \textsc{HII-CHI-mistry-Teff} sampler.}
\centering{}{\tiny\begin{tabu}{lccc}%
\hline%
Parameter&GP030321&GP101157&GP121903\\%
\hline%
$n_{e}(cm^{-3})$&$332\pm35$&$485\pm45$&$486\pm42$\\%
$T_{high}(K)$&$15200\pm200$&$13700\pm100$&$15700\pm200$\\%
$c(H\beta)$&$0.79\pm0.02$&$0.35\pm0.02$&$0.001\pm0.001$\\%
$log(U)$&$-2.56\pm0.01$&$-1.97\pm0.02$&$-2.05\pm0.01$\\%
$T_{eff}$&$74000\pm600$&$64000\pm400$&$70000\pm100$\\%
$\frac{Ar^{2+}}{H^{+}}$&$5.70\pm0.02$&$5.74\pm0.02$&$5.59\pm0.02$\\%
$\frac{Ar^{3+}}{H^{+}}$&$4.86\pm0.05$&$5.24\pm0.02$&$5.10\pm0.02$\\%
$\frac{Fe^{2+}}{H^{+}}$&$4.99\pm0.05$&$5.23\pm0.02$&$5.12\pm0.02$\\%
$\frac{N^{+}}{H^{+}}$&$6.40\pm0.06$&$6.21\pm0.11$&$6.45\pm0.04$\\%
$\frac{Ne^{2+}}{H^{+}}$&$7.19\pm0.02$&$7.19\pm0.01$&$7.02\pm0.01$\\%
$\frac{O^{+}}{H^{+}}$&$7.53\pm0.04$&$7.47\pm0.04$&$7.37\pm0.04$\\%
$\frac{O^{2+}}{H^{+}}$&$7.75\pm0.02$&$7.93\pm0.01$&$7.76\pm0.01$\\%
$\frac{S^{+}}{H^{+}}$&$5.74\pm0.02$&$5.59\pm0.02$&$5.45\pm0.02$\\%
$\frac{S^{2+}}{H^{+}}$&$6.61\pm0.05$&$6.62\pm0.04$&$6.37\pm0.04$\\%
$y^{+}$&$0.11\pm0.00$&$0.07\pm0.00$&$0.08\pm0.00$\\%
$y^{2+}$&$\num{1.9e-03}\pm\num{8e-05}$&$\num{1.3e-03}\pm\num{4e-05}$&$\num{2.2e-03}\pm\num{4e-05}$\\%
\hline%
\end{tabu}

}
\end{table}

Table \ref{tab:elemental-abundances} displays the elemental abundances computed from the output traces for the O, N, S and He ions. The methodology follows a basic approach where the measured ionic abundances are considered by most significant, and thus, they are summed up. The exception is the ionization correction factor (ICF) for the $S^{3+}$ abundance. This $ICF(S^{3+})$ was presented in \citet{fernandez_determination_2018} and it is a function of the $Ar^{2+}$ and $Ar^{3+}$ abundances. It is concluded that three GPs presented here belong to the low metallicity regime with $7.76<12+log\left(\nicefrac{O}{H}\right)<8.04$. Moreover, the galaxies display an $N/O$ excess, where GP030321 displays the largest value. This behaviour is characteristic of GPs \citep[see][]{amorin_oxygen_2010}. GP030321 galaxy also seems to have an exceedingly large helium abundance. However, as mentioned above, this element displayed a non-uniform line fitting for this galaxy. 

\subsection{Ionization source characterization} \label{radiation-character}

The results from the tailored photoionization model grids can be found in Table \ref{tab:photoIonization-param-comparison}: It displays the black body $T_{eff}$ and $log(U)$ measurements for the three GPs using the three techniques: \textsc{HII-CHI-mistry-Teff}, the Bayesian \textsc{HII-CHI-mistry-Teff} implementation and the Bayesian Direct method + \textsc{HII-CHI-mistry-Teff}. The first technique was applied using both spherical and plane parallel model geometries, while the Bayesian approaches only consider the plane-parallel geometry. The \textsc{HII-CHI-mistry-Teff} results predict a very hot ionizing radiation in all three GPs. The spherical geometry assumption consistently displays lower temperatures with $\overline{T_{eff,\, sph}}=57000\pm2800$, while the plane-parallel geometry stays at  $\overline{T_{eff,\, pp}}=61400\pm3100$. The ionization parameter values are very similar between both geometries with $\overline{log(U)}=-2.58\pm0.26$. Given the results uncertainties and taking into account the fact that real nebulae lay somewhere between these ideal geometries, it can be concluded that the grids geometry has a very weak impact on the measurements.

Carrying on with the measurements from the Bayesian \textsc{HII-CHI-mistry-Teff} implementation (column (5) in Table \ref{tab:photoIonization-param-comparison}) it can be appreciated consistent higher values for both parameters with $\overline{T_{eff}}=69915\pm369$ and $\overline{log(U)}=-2.28\pm0.01$. This behaviour was not observed on the synthetic test cases where both techniques display a good agreement. To gain some insight on the cause behind these discrepancies, the reader is advised to check the data in Table \ref{tab:photoIonization-flux-comparison}, which compares the fluxes fitted by each technique. It should be explained that the \textsc{HII-CHI-mistry-Teff} grids considers the combined flux for the $[SII]6717,6731A$ lines. This table corresponds to the measurements for GP101157. However, the corresponding tables for the rest of the sample can be found on this manuscript online support material. The percentages at each flux right-hand side quantify the difference with the observed value. It can be appreciated that all techniques display good accuracy for the collisionally excited lines. However, the $HeI$ and $HeII$ emission are poorly fitted. This pattern is repeated in the three GPs. As discussed in the previous section, the Bayesian direct method encountered large discrepancies in the $HeI$ lines for GP030321 and GP101157. However, the disagreement was also appreciated in GP121903. In this GP, the maximum discrepancy in the $HeI$ lines was below 4\% for the Bayesian direct method, while the $HeI$ fluxes predicted by the three \textsc{HII-CHI-mistry-Teff} implementations are below half the observational value. The poor fitting in these lines could explain the different results on the $T_{eff}$ and $log(U)$ measurements.

As discussed by \citet{stasinska_what_2007}, at this point we face the three challenges in photoionization modelling: Parameter space exploration, error bar characterization and model validity. Regarding the second point, the error bar characterization, the discussion in section \ref{stellar-discussion} has already illustrated this issue: Since the error propagation is lost in stellar and nebular continua computation, the impact from the underlying stellar population might be severely underestimated. While this may not be an issue for the strongest $HI$ transitions, this systematic uncertainty in the $HeI$ and $HeII$ fluxes should not be neglected. This demands the development of new techniques, which include the galaxy continua fitting in the helium chemical analysis. Regarding the third challenge, validating the ionization model for the HeII emission is still an ongoing effort both in the local \citep[see][]{shirazi_strongly_2012,senchyna_ultraviolet_2017} and high-z galaxies \citep[see][]{berg_window_2018,nanayakkara_exploring_2019}.

To approach the first issue: Parameter space exploration, we propose to combine the \textsc{HII-CHI-mistry-Teff} model with the chemical sampler in section \ref{chemical_analysis}. This approached is labelled as Bayesian direct method + \textsc{HII-CHI-mistry-Teff}. In this scheme, both model dimensions are sampled simultaneously bringing the number of fitted parameters to 16. To the best authors knowledge, this is the largest chemical model for the fitting of the recombination and collisionally excited fluxes. Moreover, it also represents the first attempt to simultaneously include ionization parameters in a direct method fitting. The closest analogue to this methodology (in addition to \textsc{HII-CHI-mistry}) is the BOND (Bayesian Oxygen and Nitrogen Determination) algorithm by \citet{vale_asari_bond:_2016}. The program derives the oxygen and nitrogen abundances  via the strong line method paradigm. However, their code also considers the relative $HeI$ fluxes to anchor the hardness of the radiation field. Checking column (6) in Table \ref{tab:photoIonization-param-comparison}, we can appreciate that the combined technique results in $T_{eff}$ and $log(U)$ measurements similar to those from the Bayesian \textsc{HII-CHI-mistry-Teff} algorithm. Moreover, the flux comparison in Table \ref{tab:photoIonization-flux-comparison} leads to the conclusion that once both techniques are combined the observational data is better represented for both recombination and collisionally excited lines. Nonetheless, a careful analysis on these results points towards discrepancies between the theoretical model and the observational data:

\begin{itemize}
    \item The difference between the fluxes predicted by the Bayesian direct method + \textsc{HII-CHI-mistry-Teff} algorithm is still higher than in the original Bayesian direct method. It is thus necessary to confirm that this is not a mathematical issue, since as the number of dimensions increases, so is the quality of the fitting is expected to rise. For example, in the Bayesian direct method fittings, there is usually a very good match for input spectra where an ionic species is only responsible for a unique emission line. In this case, the algorithm has a simple job to find the abundance, which best simulates the observed flux; even as the shared parameters vary. This same pattern may be happening in the photoionization model sampling. Consequently, several lines should be included for every species, whenever possible. This conclusion also applies to both the input spectra and the considered theoretical photoionization models.
    
    \item In the case of GP101157, the fitting results in double peak posteriors for the model parameters. This is because the sampling is struggling to find the model solution at two different coordinates. However, looking at the emission flux posteriors, only the $[OIII]$ and $HI$ transitions show a clear double peak shape. This means that in the current model the $\nicefrac{O}{H}$ and $c(H\beta)$ sampling have the largest impact. Since this phenomenon is only appreciated in the case of GP101157, it may be argued that this galaxy displays an abnormal behaviour (which could also explain the divergence on the $HeI$ fluxes). Nonetheless, it is necessary to check if the oxygen abundance has a disproportionate weight on the grid interpolation. Since it is well known that in the very low metallicity region the chemical enrichment is linear  \citep[see][]{peimbert_chemical_1969,peimbert_chemical_2000}, the photoionization grids could be interpolated as function of the total metallicity. Afterwards, the metallicity could be expressed as a linear function of all the fitted abundances. This way, we could guarantee that all species contribute to the abundance interpolation.
    
    \item Both the $T_{eff}$ and $log(U)$ measurements do not change dramatically in the Bayesian direct method + \textsc{HII-CHI-mistry-Teff} fitting even though the $HeI$ and $HeII$ fluxes are closer to the observational values. These lead to two possible conclusions: The first one is that the model sampling is failing to establish the GPs ionizing radiation strength from the helium photons. To confirm this is not an issue, the photoionization model should be tested with several diagnostics for the same photoionization grids. For example, the $T_{eff}$ could be characterised as a function of the softness parameter \citep[see][]{vilchez_determination_1988}. Similarly, the ionization parameter could be expressed as a function of the $\nicefrac{O^{+}}{O^{2+}}$ or $\nicefrac{S^{+}}{S^{2+}}$ ratios. The second scenario, however, is that $HeII$ fluxes cannot be fully explained by the photoionization model. Indeed, radiation only driven by stellar atmospheres usually fails to explain these ions origin \citep[see][]{stasinska_modeling_2003,kehrig_extended_2015,kehrig_extended_2018}. This result suggests that additional phenomena (shocks, photon leakage, nearly metal-free hot stars, X-ray binaries...) are necessary to account for the helium photons excess in star-forming galaxies. Recently, the models from \citet{simmonds_can_2021} have shown that ultra-luminous X-ray sources may provide the necessary ionizing radiation for the observed HeII fluxes. The next step in this line of work, is to enhance the current fitting algorithm for such grids. Afterwards, it will be possible to consistently compare their accuracy for all the available lines.

\end{itemize}

Table \ref{tab:fit_results_large} displays the complete results from the Bayesian direct method + \textsc{HII-CHI-mistry-Teff} fitting. We find that the new model predicts slightly higher densities and lower temperatures. The largest discrepancies are found on the electron temperature for GP030321 and GP101157, which are significantly lower in the new fitting $(\Delta T_{high}\approx 1000K)$. This has the expected effect on the metal abundances. Taking into consideration the points previously discussed, this result is caused by the fitting of the $[OII]3726\AA,3720\AA$ and $[OIII]5007\AA$ lines as the sampling is dominated by the photoionization model grids. However, in the case of GP12903, the galaxy with the best match between the observed fluxes and the theoretical ones, both model techniques provide consistent results. These results support the combination of both techniques. At the current time, however, we shall favour the results from the isolated Bayesian direct method and the \textsc{HII-CHI-mistry-Teff} algorithm with spherical geometry. To further explore the physical validity of this technique, we shall work on larger photoionization models, which include all the emission lines observed. Needless to say, such large simulations must be accompanied by many test cases to diagnose the previous issues.

\section{Summary and Conclusions} \label{conclusion}

In this paper, we present deep long-slit optical spectra from the OSIRIS instrument at the 10-m GTC for three Green Pea galaxies, GP030321,GP101157 and GP121903. In addition to the application of state-of-the-art tools from the literature, a goal of this study is to apply novel algorithms for the ionised spectrum fitting. The GP sample treatment has its dust extinction quantified, its nebular and stellar continua fit and their chemical compositions characterised. Additionally, we apply the same methods for the reanalysis of another three GPs from \citetalias{amorin_star_2012}, which were observed using the same instrument and setups: GP004054, GP113303 and GP232539. The new galaxies redshift stays in the $0.144 < z < 0.196$ interval. The main conclusions from this work are:

The galaxy sample is characterized by strong emission lines $(EqW([OIII])~2165-2926\text{\AA}$; $EqW(H\beta)\approx595-739\text{\AA})$ and faint continua. Our deep spectra allow us to detect faint lines, such as HeII and several metal auroral lines. Unlike in the three GPs presented in \citetalias{amorin_star_2012}, the new sample continua do not display stellar absorptions or WR bumps. 

Two approaches are considered in GP continua analysis. The first one consists in the nebular continua calculation from first principles using the chemical properties measured from the gas. Afterwards, a stellar population synthesis is applied on the observed spectrum without the gas component, using \textsc{Starlight}. The output is used to correct the recombination lines from underlying stellar absorption before repeating the process. The second methodology consists in the application of self-consistent population synthesis code \textsc{FADO}. In addition to the inbuilt nebular continuum computation, this algorithm was used to fit the largest SSP library ever considered for GPs: a total 1098 stellar spectra grid with fine time steps. Both techniques display consistent results despite the different methodologies and inputs. However, the \textsc{FADO} implementation provides a simpler and more robust workflow. The results are the ones used for the SFH analysis, while those from the iterative approach are used in the chemical study.

In GP101157 and GP121903, the new methodology provides very good agreement with the stellar masses presented in \citetalias{amorin_star_2012}. In the case of GP004054, however, the stellar mass is a $13\%$ lower than the one measured in \citetalias{amorin_star_2012}. This is because the older stellar fraction is not fitted. In general, however, the new approach succeeds at fitting stellar populations from a few hundred thousand years old to several Gyr. The light fraction is dominated by the $6.5 < \Delta log(t) < 7.5\, yr$ age range, while younger stars only account for a few percentage points. Older stellar populations dominate the mass and they formed several Gyr ago. Unlike in the younger systems, old stars do not share a common distribution for the mass and light fraction histograms. Instead, a single SSP dominates the light fraction, with a value of the same order as the luminosity from the younger stars. This points towards an incomplete characterisation from the gas contribution: either an underestimation in our nebular continuum calculation, or an overestimation on the older stars SSP synthesis due to an underestimated contribution from the ionizing stellar population. The second scenario would require unaccounted phenomena such as photon leakage or the presence of almost pristine and very massive stars.

The algorithm \textsc{HII-CHI-mistry-Teff} is applied to measure the $T_eff$ and $log(U)$ from the ionizing cluster. Using the plane-parallel geometry and $30 - 90\,KT$ black body grids the model predicts very hard ionizing sources for the GP sample with $\overline{T_{eff,\, pp}}=61431\pm3076$ and $\overline{log(U)}=-1.98\pm0.24$. Additionally, using the neural networks grid interpolator from the library \textsc{exoplanet}, we propose a Bayesian implementation for \textsc{HII-CHI-mistry-Teff}. In synthetic test cases, this version provides the same accuracy for a much faster treatment. Once applied to the GPs sample, a harder ionization source is predicted with $\overline{T_{eff,\, pp}}=69915\pm369$ and $\overline{log(U)}=-2.28\pm0.24$. Both approaches underestimate the observed $HeI$ and $HeII$ fluxes. This suggest the existence of additionally mechanisms enhancing the helium transition photons.

Using the tools mentioned above, the Bayesian direct method sampler from \citet{fernandez_bayesian_2019} has been enhanced to interpolate emissivity grids during the sampling process. Additionally, it has been adapted to include other ionic species. In the current GP sample, the maximum number of dimensions is 14. The results show a low metallicity regime with high ionization temperatures, $T_{e,[OIII]} > 14000\,K$. As a final exercise, this model was combined with Bayesian \textsc{HII-CHI-mistry-Teff} algorithm. This represents the first attempt to fully solve the direct method parameter space alongside photoionization grids parameter space. The output $T_{eff}$ and $log(U)$ values agree with those from the Bayesian \textsc{HII-CHI-mistry-Teff} implementation but the measured $T_{[OIII]}$ is lower than in the original Bayesian direct method for two of the three galaxies. The predicted emission fluxes also display a slightly larger discrepancy with the observed fluxes compared with the original Bayesian direct method. At this point, further work is necessary to find the best design to include tailored photoionization grids in the chemical model fitting.
This study represents an ongoing effort to improve our understanding on the SFH and the chemical evolution of star-forming systems. These conclusions  emphasize the importance of combining deep, high S/N spectra from large telescopes with sophisticated algorithms and models. The insights gained lead to the conclusion that the default models for star forming regions and gas emission fail to explain some biases encountered in GPs. Fortunately, the novel methodologies proposed are not limited to the size of input data. This will make possible the solution of such complex models.

\section*{Acknowledgements}
We are indebted to an anonymous referee, whose detailed comments and encouragement contributed to the improvement of the paper.
This work was funded by the FONDECYT Postdoc 2020 project 3200340 from the National Agency for Research and Development (ANID). R.A. acknowledges support from ANID Fondecyt Regular 1202007. P.P. acknowledges support through FCT grants UID/FIS/04434/2019, UIDB/04434/2020, UIDP/04434/2020 and the project "Identifying the Earliest Supermassive Black Holes with ALMA (IdEaS with ALMA)" (PTDC/FIS-AST/29245/2017). EPM, CK, and JVM acknowledge financial support from the State Agency for Researchof the Spanish MCIU through the “Center of Excellence SeveroOchoa” award to the Instituto de Astrofisica de Andalucia (SEV-2017-0709). 

\section*{Data Availability}

The authors make available with this manuscript the following data on the corresponding MNRAS supplementary material section: 1) The emission line fluxes in machine readable format. 2) The full results from the Bayesian fittings both in graphical and tabulated format. 3) The complete infographics from the FADO synthesis in portable document format (PDF). The observations files can be downloaded from the GTC virutal observatory at https://gtc.sdc.cab.inta-csic.es/gtc/jsp/searchform.jsp. The program ID is GTC63-10B.


\bibliographystyle{mnras}
\bibliography{references_zotero.bib}



\appendix

\section{FADO results}

This appendix includes the results from the FADO fitting for the sample described in this manuscript, as well as, the one in \cite{amorin_star_2012}. These plots share the same format as the one in Fig.\ref{fig:fado-infograph}.

\begin{figure*}
\includegraphics[width=1.0\textwidth]{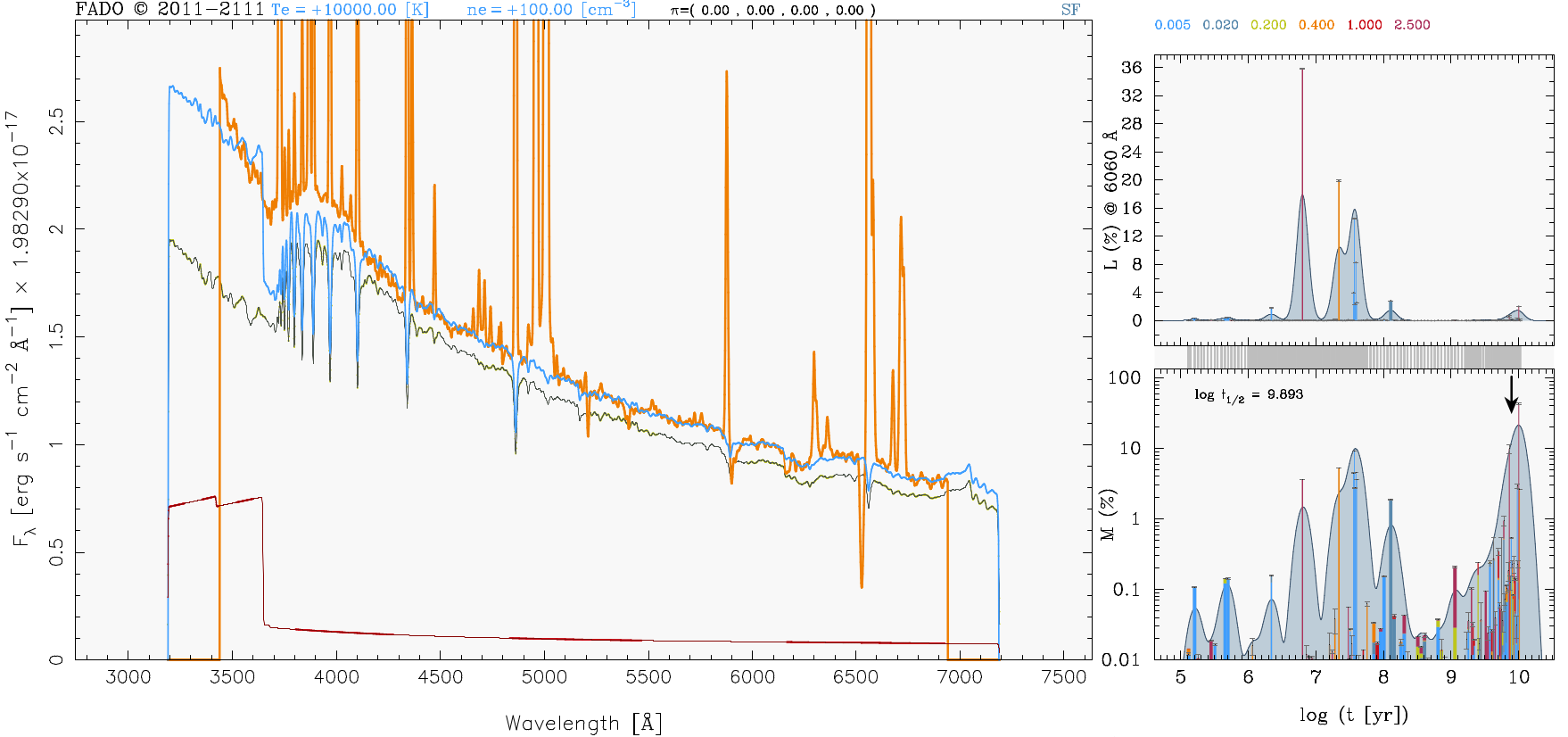}
\caption{\label{fig:fado-infographA}\textsc{FADO} continua fit for the galaxy GP030321 (orange). The best fitting SED (blue) includes the stellar (black) and nebular (red) components. }
\end{figure*}

\begin{figure*}
\includegraphics[width=1.0\textwidth]{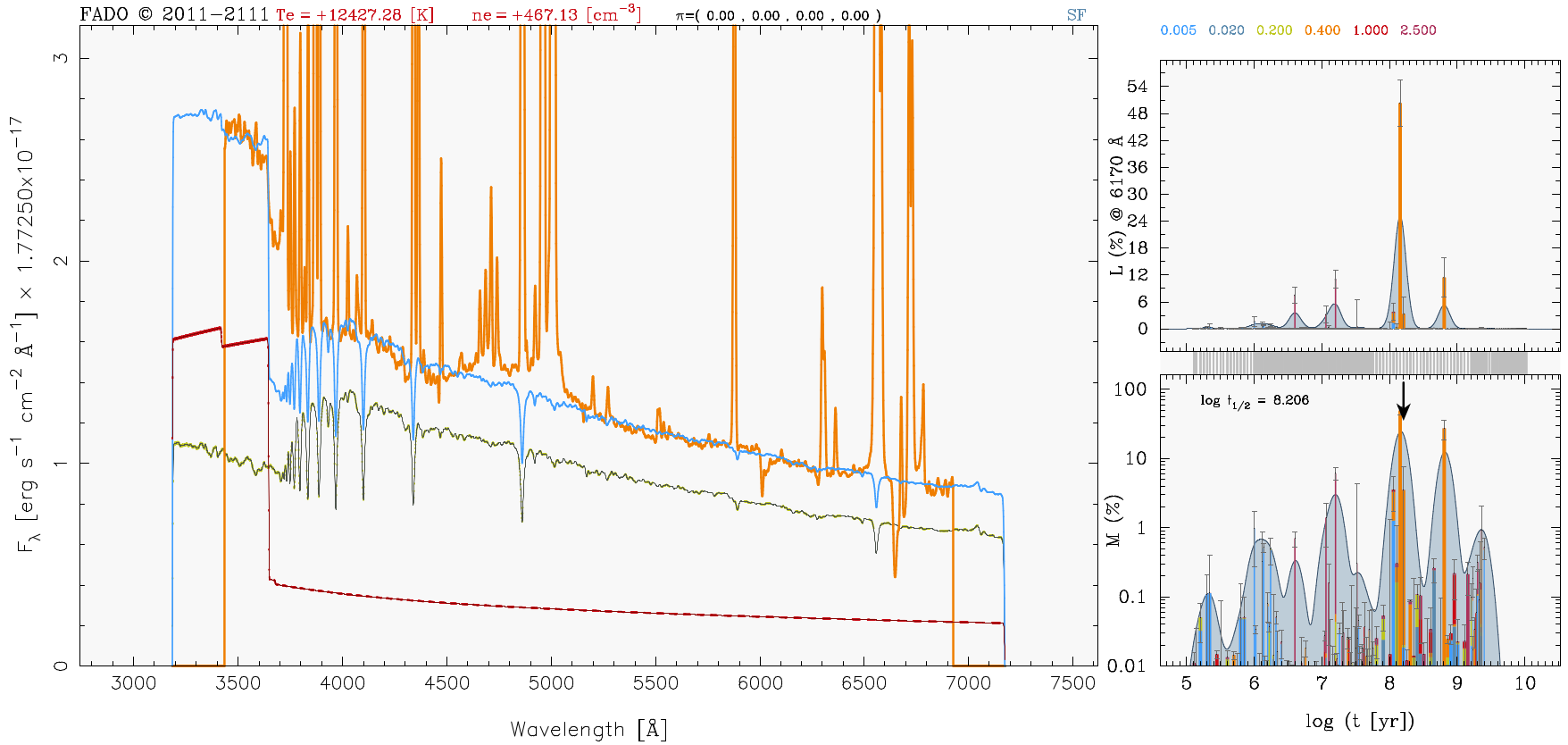}
\caption{\label{fig:fado-infographB}\textsc{FADO} continua fit for the galaxy GP101157 (orange). The best fitting SED (blue) includes the stellar (black) and nebular (red) components. }
\end{figure*}

\begin{figure*}
\includegraphics[width=1.0\textwidth]{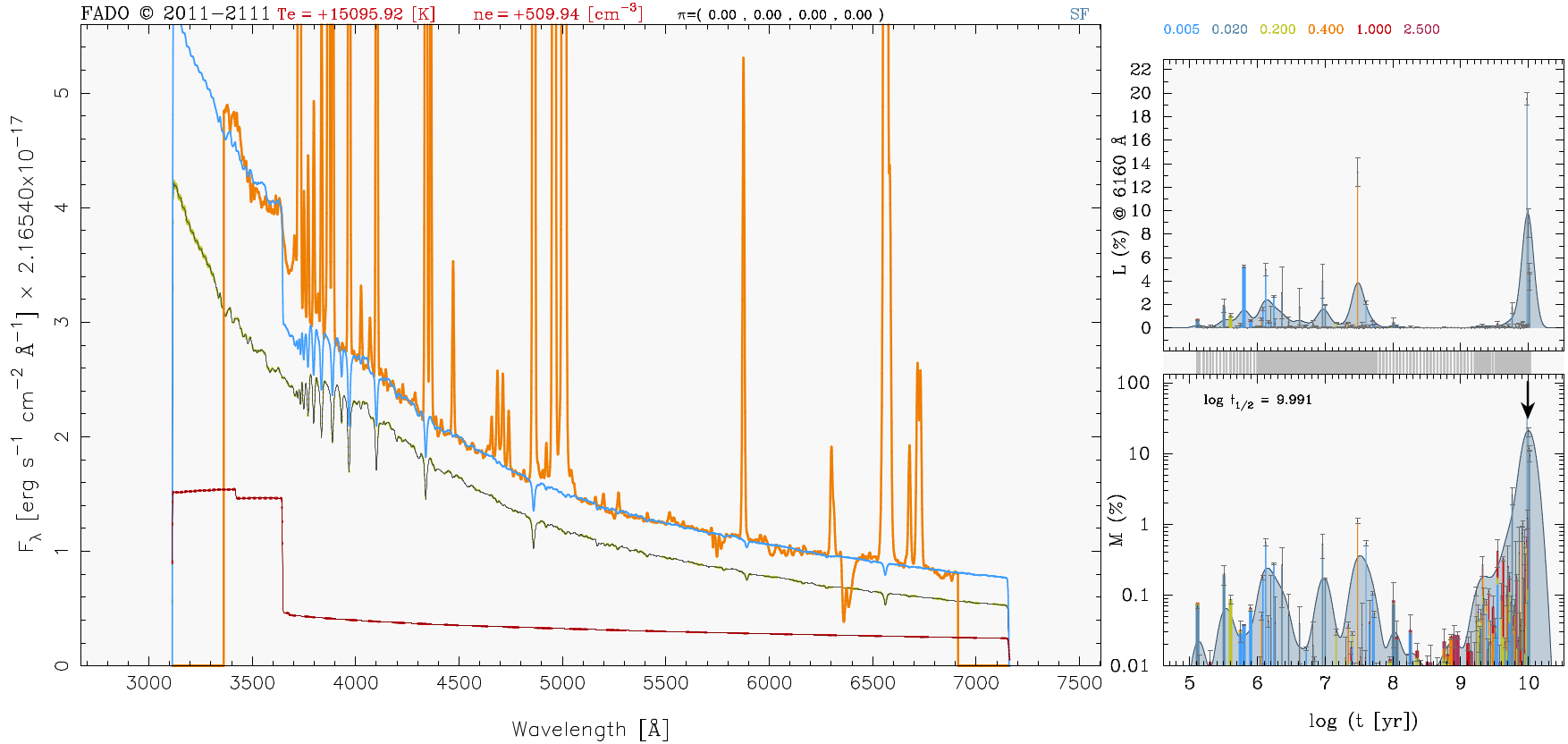}
\caption{\label{fig:fado-infographC}\textsc{FADO} continua fit for the galaxy GP121903 (orange). The best fitting SED (blue) includes the stellar (black) and nebular (red) components. }
\end{figure*}

\begin{figure*}
\includegraphics[width=1.0\textwidth]{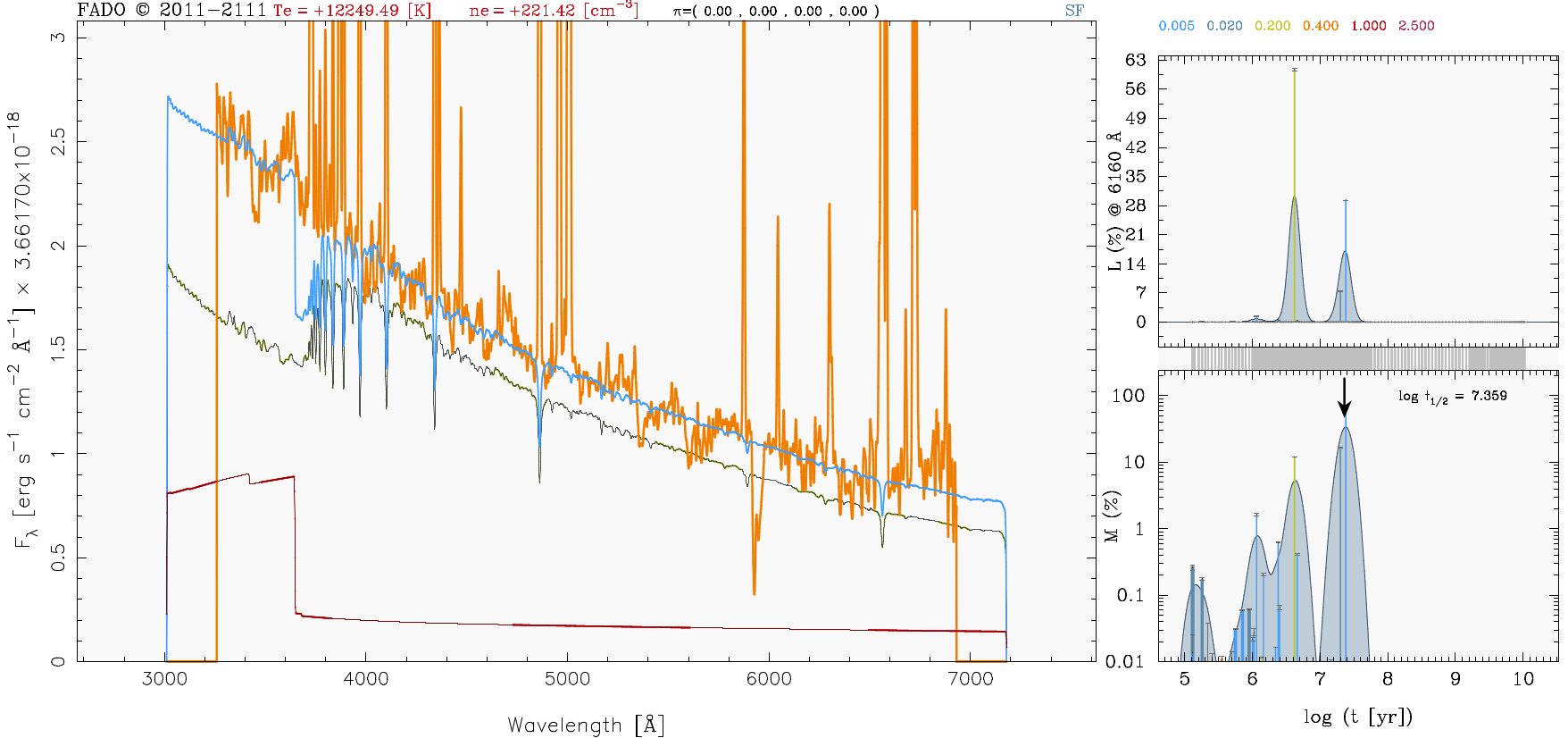}
\caption{\label{fig:fado-infographD}\textsc{FADO} continua fit for the galaxy GP004054 (orange). The best fitting SED (blue) includes the stellar (black) and nebular (red) components. }
\end{figure*}

\begin{figure*}
\includegraphics[width=1.0\textwidth]{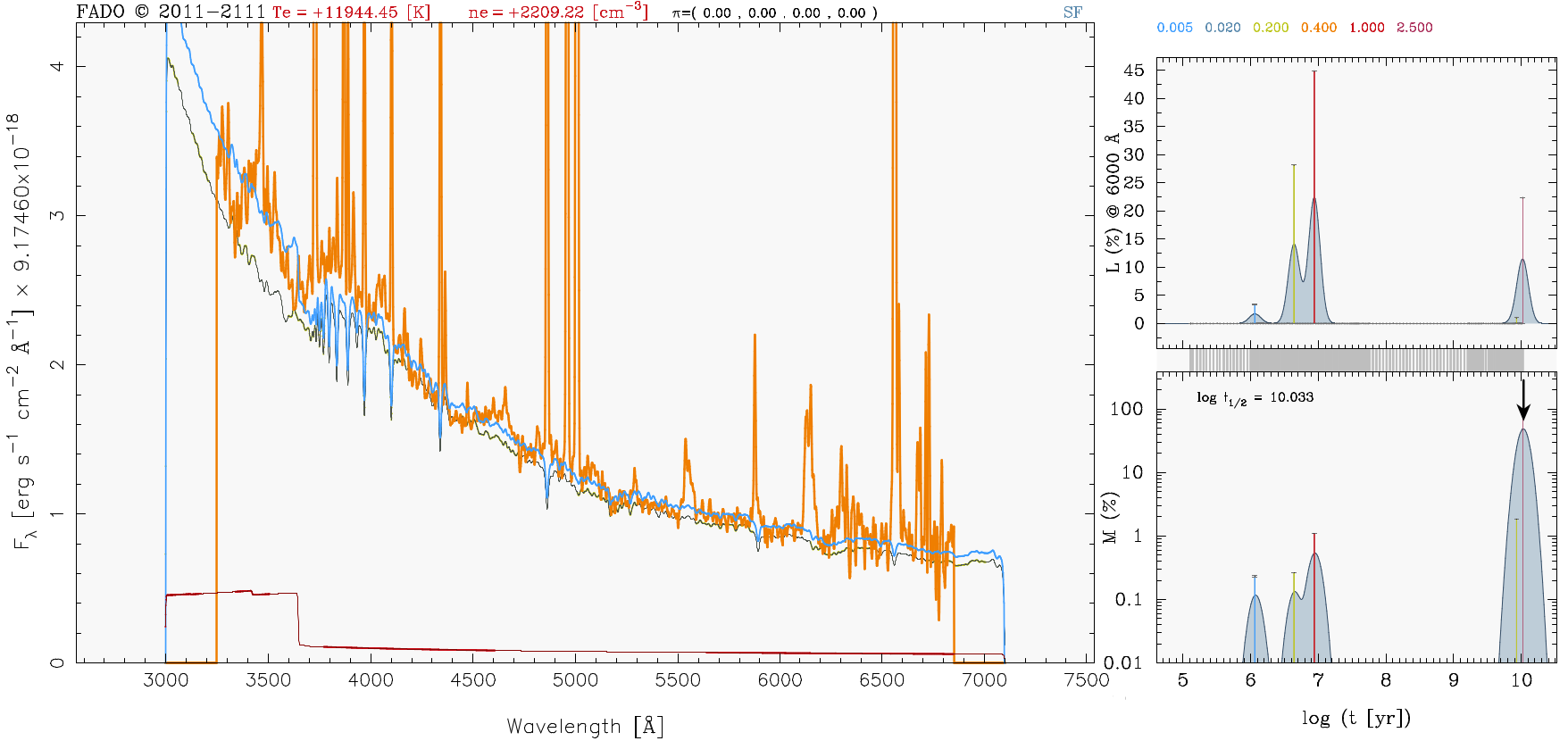}
\caption{\label{fig:fado-infographE}\textsc{FADO} continua fit for the galaxy GP113303 (orange). The best fitting SED (blue) includes the stellar (black) and nebular (red) components. }
\end{figure*}


\bsp	
\label{lastpage}
\end{document}